%% file: main.tex
\documentclass[notitlepage]{article}
\usepackage{graphicx} % Required for inserting images
\usepackage{authblk}
\usepackage{amsmath, booktabs}
\usepackage[backend=biber, style=numeric, sorting=none, doi=false,isbn=false,url=false,eprint=false]{biblatex}
\usepackage{algorithm, algpseudocode}
\usepackage{subcaption}

\title{Simulation-based Criticism of Complex Biological Models with Neural Evidence Estimation}
\author[1]{Prayag Chatha}
\author[2]{Fan Bu}
\author[1]{Jon Zelner}
\affil[1]{Department of Epidemiology, University of Michigan, Ann Arbor, Michigan, United States of America}
\affil[2]{Department of Biostatistics, University of Michigan, Ann Arbor, Michigan, United States of America}
\date{}

\newcommand{\bth}{\boldsymbol{\theta}} \newcommand{\bx}{\mathbf{x}}
\newcommand{\mdl}{\mathcal{M}}
\DeclareMathOperator*{\argmin}{arg\,min}

\newcommand{\bX}{\mathbf{X}}
\newcommand{\bY}{\mathbf{Y}}
\algrenewcommand\algorithmicrequire{\textbf{Input:}}
\algrenewcommand\algorithmicensure{\textbf{Output:}}
\newcommand{\by}{\mathbf{y}}
\newcommand{\bbeta}{\boldsymbol{\beta}}
\newcommand{\bz}{\mathbf{z}}

\begin{document}

\maketitle

\begin{abstract}
    In statistical modeling, the likelihood is the basis for both inference and criticism, which constitute the two stages of data analysis. However, many realistic scientific models lack computationally tractable likelihoods. Increasingly, these are implemented as computer simulations. New deep learning-based methods for simulation-based inference (SBI) have demonstrated the ability to efficiently and accurately calibrate complex simulation models. However, in this setting, the standard tools for comparing models and assessing their potential limitations are often unavailable. We propose a new method for simulation-based model criticism, Neural Evidence Estimation. NEE trains a normalizing flow on simulated data to learn an estimate of the marginal likelihood (a.k.a. evidence). These estimates can be used to diagnose model misspecification and perform model comparison without evaluating (or approximating) a model's likelihood. Compared to related techniques designed for either of these tasks, NEE is quick, robust, and modular. We demonstrate our methodology through applications to infectious disease modeling. Here, model criticism is difficult yet imperative: infectious disease modelers often face a choice among several plausible models featuring intractable likelihoods. We first validate NEE's ability to discern incorrect transmission models through simulation experiments. We then perform data analysis of a historic multi-wave influenza epidemic, where we use SBI and NEE to rigorously compare several competing mechanistic theories of reinfection. Overall, NEE provides a unifying framework for simulation-based criticism, advances efficient multi-model data analysis, and helps improve the reliability of SBI.
\end{abstract}

\clearpage
% Stochastic Simulation Model Comparison with Neural Evidence Estimation
% NEE: Comparison and Criticism of Stochastic Simulation Models
% Criticism of Stochastic Simulation Models in Epidemiology with Neural Evidence Estimation

\section*{Author Summary}

Scientific data analysis can be conceptualized as an iterative process of inference and criticism. In the inference stage, scientists fit a theory---ideally, a model making quantitative predictions---to observed data in order to derive generalizable insights. In the criticism stage, scientists use data to identify potential limitations of the model and compare it against other candidates, which drives the development of new theories. In classical statistics, the foundation of inference and criticism is a model's likelihood, a function that relates the probability of observed data to unobserved parameters. However, computing the likelihood may be prohibitively slow for complex statistical models seen in scientific applications. These types of models tend to be implemented as computer simulations. A new set of methods known as simulation-based inference (SBI) use neural networks trained on simulated data to efficiently fit such models to observations, but the standard toolkit for model criticism is often inadequate or unavailable in this setting.

We propose a method for criticism of complex simulation models that avoids likelihood evaluation, Neural Evidence Estimation (NEE). NEE trains a flexible neural network on simulations to learn a quantity known as the model evidence, which can be thought of as the probability of an observation being generated by a given, unfitted model. Depending on the task at hand, the estimate of evidence can then be used to diagnose when a model is a bad description of reality, select between multiple candidate models, or generate an ensemble average of multiple models. We apply NEE to problems in transmission modeling of infectious diseases. These models often lack easily computable likelihoods, yet model comparison plays an important role in epidemiology. We show that NEE offers an efficient, robust, and modular solution for the problem of simulation-based model criticism. We argue that applying critical scrutiny to simulation models helps ensure the reliability of SBI.

% what's the selling point for NEE?
% can wrap up with a homily about the importance of taking a critical attitude to models, especially when using deep learning to make predictions based on these models. 

\clearpage

% begin with: complex simulation models are widely used yadda yadda but are poorly suited for conventional statistical inference.

% In infectious disease epidemiology, thanks to the complexity of systems-level interactions and limited data, there are typically numerous mechanistic models that can explain key patterns in a given dataset. Statistical model criticism aims to quantify the uncertainty inherent in data modeling to guide decision making and test hypotheses. However, cl-assical techniques for model criticism are ill-suited for complex simulations of infectious disease, which have become increasingly prominent in epidemiology.

% We propose a new method for criticism of transmission simulation models, Neural Evidence Estimation (NEE). NEE trains neural networks on simulated data (e.g., trajectories of an epidemic's incidence) to approximate the likelihood of a model with respect to observed data, or \textit{evidence}. We then use the evidence to select between multiple models and test possible model misspecification. 

% We apply NEE to simulation experiments and a case study of modeling a multiple wave influenza epidemic. We highlight a computationally efficient yet conceptually simple methodology for multi-model analysis of epidemiological data. Our work demonstrates how model criticism can bolster the reliability of neural techniques for simulation-based statistical inference.

% \clearpage

\section{Introduction}

% cite: Gelman, A. and Shalizi, C. R. (2013). Philosophy and the practice of Bayesian statistics. British
% Journal of Mathematical and Statistical Psychology, 66(1):8–38.
% what is model criticism? why do we care?
Model criticism has long been recognized as an essential component of statistical data analysis \cite{box1976science, gelman2013philosophy}. Data modeling advances the development of scientific theories through an iterative process of inference and criticism: scientists fit a model to observations in order to draw generalizable conclusions, and then they use observations to diagnose possible shortcomings of the model and compare the model to other candidates \cite{mackay1992bayesian}. In classical statistics, the likelihood---the probability of some observed data conditional on a specific set of model inputs---is the basis for both parameter estimation and model comparison, yet for many realistic, complex models, computation of the likelihood may be infeasible. Thus, whenever model fitting is difficult, then model criticism is difficult too.
% a contradiction between the data and the model is often an indication that the model's underlying assumptions are flawed, which often drives scientific discovery.

In scientific applications, such data generating processes are often implemented as computer simulations rather than strict probabilistic models. Formally, we define a simulation to be a program accepting parameters $\bth$ as input that stochastically generate data $\bx$ as output, thereby implicitly defining the model likelihood, $p(\bx \mid \bth).$ \textit{Simulation-based inference} (SBI) refers to an emergent family of statistical techniques in which unknown parameters $\bth$ for an observed dataset $\bx_o$ are inferred through analysis of many simulated datasets $\bx~\sim p(\bx \mid \bth)$ \cite{cranmer_frontier_2020}. Notable among these is Neural Posterior Estimation (NPE), where a neural network trained on simulated data learns an conditional density estimator $q(\bth \mid \bx)$ approximating the model posterior $p(\bth \mid \bx)$. NPE avoids direct evaluation of the likelihood, so it provides a computationally efficient solution for calibrating models with intractable likelihoods \cite{papamakarios_fast_2018}. Such computational advances have facilitated inference with complex simulation models, yet the standard methodologies for model comparison and diagnosis are unavailable in this setting. 

In this work, we propose a deep learning-driven technique for quantitative criticism of stochastic simulation models: Neural Evidence Estimation (NEE). Analogously to NPE, NEE uses normalizing flows---a flexible neural network architecture that has achieved state-of-the-art results in estimating complex probability distributions \cite{papamakarios2021normalizing}---trained on simulated data to estimate the \textit{marginal likelihood},
$p_k(\bx) = \int_{\bth} p_k(\bx, \bth) d\bth$. In Bayesian inference, the marginal likelihood is a natural measure of a given model's evidence with respect to a dataset \cite{mackay1992bayesian}. We present a theoretical framework for simulation-based criticism that unifies diagnosis of model misspecification and model comparison, two intrinsically related problems that have until now been studied separately from one another. In our experiments, we show how NEE can be used for both frequentist hypothesis testing of model misspecification and Bayesian model selection and averaging while circumventing evaluation (or approximation) of model likelihoods. Compared to the available tools for simulation-based criticism, NEE is a conceptually simple, computationally efficient, one-stage estimation procedure and is modular with respect to the inference procedure used and the simulators' design.

In this work, we show NEE's utility for statistical criticism by application to problems in infectious disease modeling. Since they offer mechanistic interpretability of systems and quantification of uncertainty in outcomes, stochastic compartmental models have become a popular tool in epidemiology (and related disciplines, e.g. ecology) \cite{andersson2012stochastic, black2012stochastic}. Nonetheless, fitting these models to observations can be quite difficult: they are typically nonlinear dynamical systems featuring many latent variables corresponding to unobserved events \cite{breto2009time}, so their likelihoods are often computationally intractable for even modestly sized populations, which has motivated a variety of statistical calibration techniques \cite{he2009plug, McKinley2018Approximate, endo2019introduction}. There is additionally growing interest in using deep learning methods, including NPE, to help fit large scale infectious disease models \cite{la2020epidemiological, kosma2023neural, ahmad2025can, pinotti2026simulation}. While these neural approaches may be computationally fast, it is can be difficult to rigorously evaluate the inferences they draw. In particular, NPE's reliability depends strongly on the versimilitude of the simulator used to generate its training samples \cite{cannon2022investigating, ward_robust_2022}. We use examples from infectious disease modeling to argue that NEE helps improve the robustness of NPE's inference with little computational overhead.

The complex, multi-layered nature of the systems studied in epidemiology, ecology, and related life sciences makes principled model criticism especially important \cite{hooten_guide_2015, thoren2024model, zuchowski2024rarely}. Mechanistic models of organic, evolutionarily adaptive processes entail significant abstractions compared to the physical processes studied in, say, cosmology or particle physics, where one can expect a cleaner fit between observation and mathematical analysis. Stochastic compartmental models represent, in an idealized manner, pathogen-host interactions at scales ranging from the cellular to the social; measurement of epidemiological outcomes is difficult and is seldom near-complete \cite{metcalf2017opportunities, crepey2022challenges}. For these reasons, when embarking on a modeling analysis, infectious disease modelers often face a difficult choice between several models that present competing---yet, at first glance, equally plausible---explanations regarding a given dataset. These models make distinct assumptions and often encode competing theories, e.g. heterogeneous versus random mixing. Since multi-model analysis can be expected to offer more robust predictions and insights than inference with a single model, techniques for model selection and model ensembling are employed widely in infectious disease epidemiology \cite{cramer2022evaluation, sherratt2023predictive, craddock2025bayesian}. However, the existing toolkit for model comparison in this domain is often inadequate when models increase in scale.

Through simulation experiments and a real-world case study of reinfection in an influenza outbreak, we demonstrate a simulation-based workflow for inference and criticism featuring NEE. We argue that NEE is a practical solution to the problem of rigorously comparing stochastic compartmental models. While our work focuses on applications to infectious disease modeling, our approach is generalizable to other branches of scientific simulation modeling.

\section{Methods}
\label{sec:methods}

\subsection{Simulation-based Inference}
\label{sec:sbi}

% We sample from the posterior predictive distribution by 

% A model's posterior predictive density is
% \begin{equation}
%     p(\bx \mid \bx_o) = p(\bx \mid \bth) \cdot p(\bth \mid \bx_o)
% \end{equation};
% we can sample from it by sampling parameters from the posterior and plugging them into the simulator.

Let $\bx_o$ denote an observed dataset that depends on an unknown set of parameters $\bth$ via a likelihood density 
$p(\bx_o \mid \bth)$. We assume that $\bth$ follows a prior distribution $p(\bth).$ The primary aim of Bayesian inference is to compute the posterior distribution of $\bth$ conditional on $\bx_o.$ Using Bayes' rule, we can express this conditional distribution as
\begin{equation}
\label{eq:posterior}
    p(\bth \mid \bx_o) = \frac{p(\bx_o \mid \bth) p(\bth)}{p(\bx_o)}.
\end{equation}
The posterior $p(\bth \mid \bx_o)$ provides an estimate of the unknown parameters $\bth$ with uncertainty quantification. Beyond inference, the posterior can be used for predicting unobserved realizations $\bx$ based on the observed $\bx_o$ for a fitted model through the posterior predictive distribution,
\begin{equation}
\label{eq:ppd}
    p(\bx \mid \bx_o) = p(\bx \mid \bth) p(\bth \mid \bx_o).
\end{equation}

The numerator in Equation~\ref{eq:posterior} is the joint distribution of the prior and the likelihood, known as the \textit{model}. The denominator is known as the evidence or marginal likelihood, which is hard to solve analytically for all but the simplest models, motivating the many techniques of Bayesian computation. Conventional Bayesian inference, e.g. Markov Chain Monte Carlo (MCMC), relies repeated evaluation of an analytic likelihood, so it is ill-suited for simulation models with intractable likelihoods. Nonetheless, sampling from simulators (as the name would suggest) is generally straightforward even when computing the likelihood is not \cite{cranmer_frontier_2020}.

Neural Posterior Estimation (NPE), which is the oldest form of neural SBI, approximates the posterior by training a flexible neural conditional density estimator $q_\phi (\cdot \mid \cdot)$ on many parameter-data pairs $(\bth_s, \bx_s)$ sampled from the model $p(\bth, \bx)$. Here, $\phi$ denotes the trainable weights (parameters) defining an encoder network that maps simulated data $\bx$ to the conditional density $q_\phi(\bth \mid \bx),$ effectively inverting the forward simulation model \cite{papamakarios_fast_2018}. During training, NPE maximizes the objective function 
\begin{equation}
    \mathbf{E}_{p(\bth, \bx)} [\log q_{\phi} (\bth \mid \bx)].
\end{equation} After training, the observed data $\bx_o$ is plugged into the density estimator to obtain a surrogate posterior approximation, $\hat p(\bth \mid \bx_o) = q_{\phi}(\bth \mid \bx_o)$. 

NPE can attain improved sample efficiency compared to Monte Carlo sampling-based Bayesian methods, such as MCMC and ABC, because it performs \textit{amortized inference}: NPE learns the general posterior $p(\bth \mid \bx)$ over many simulated $\bx$ to arrive at a target posterior $p(\bth \mid \bx_o)$. NPE leverages the strength of neural networks at interpolating high-dimensional data to predict the posterior from relatively scarce training examples. However, the validity of NPE and other neural SBI techniques rests on the adequacy of the simulator with respect to the real data generating process \cite{ward_robust_2022}. Should simulations fail to resemble the observation(s) $\bx_o$, amortized inference amounts to \textit{out-of-distribution prediction}, a notorious vulnerability of deep learning. Thanks to their flexibility, neural networks are prone to overfitting their training distribution, so an encoder network trained on bad simulations will yield poor predictions on real data. Succinctly, SBI is governed by the famous law of computation, ``garbage in, garbage out.''

So-called \textit{model misspecification} is an ubiquitous hazard in scientific applications, since simulators are idealized representations of reality. While misspecification can be problematic for likelihood-based Bayesian inference, it has been shown that the MCMC  will asymptotically converge on a wrong-yet-interpretable estimate, namely the point in parameter space that minimizes the Kullback-Liebler (KL) divergence between the true model and the misspecified model \cite{kleijn2012bernstein}. A similar result has been obtained for the simplest forms of ABC \cite{frazier2020model}, but as of yet, there is no such theoretical guarantee for neural SBI techniques.

We formalize the problem of model specification as it pertains to NPE, following \textcite{ward_robust_2022}. Suppose $\bx_o$ comes from a true, unknown distribution $p^*$. The simulator $p(\bx \mid \bth)$ is misspecified with respect to $\bx_o$ if $p^* \notin \{p(\bx \mid \bth); \bth \in \boldsymbol{\Theta}\}$. Of course, outside of synthetic experiments, we have no knowledge of $p^*$, but we can measure misspecification through the proxy of marginal likelihood. When $p(\bx_o)$ is close to zero, then few of the sampled simulations $\bx \sim \int_{\bth}p(\bx_o \mid \bth)p(\bth)$ will resemble $\bx_o$, no matter how large the simulation budget. When $\bx_o$ is entirely outside the support of $p(\bx)$, then NPE amounts to attempted extrapolation. In either scenario, NPE may yield an inaccurate posterior estimate \cite{cannon2022investigating}.

Misspecification in SBI can arise from either a bad prior---i.e. a prior whose support fails to cover the true posterior---or a simulator that fails to generate realistic data. We focus on addressing the latter problem in this article. In applied analysis, simulators are usually complex computer programs involving many design choices, whereas priors are often taken to be weakly informative, simple distributions (e.g. normal or uniform). While more consideration may go into specifying a likelihood, prior choice matters a great deal when the sample size is small, there is weak likelihood signal in the data, or when there are many unknown parameters. For a theoretical perspective on prior specification, we refer readers to \textcite{gelman2017prior}. 

\subsection{Normalizing Flows}

Normalizing flows are a family of neural network architectures that compose several simple, invertible functions into a ``flow'' that can map data from a complex distribution into a simple distribution, such as a multivariate normal. Inverting this flow allows for approximate sampling from the unknown target distribution as well as density estimation using the transformation of variables theorem \cite{papamakarios2021normalizing}. Thanks to their flexibility, they are considered state-of-the-art for marginal and conditional density estimation. They are commonly used in SBI to learn an approximation of the posterior or likelihood densities from simulated data.

In this work, we use Masked Autoregressive Flows (MAF, \cite{papamakarios2017masked}, as implemented in the \textbf{Zuko} Python library, for the density estimators powering both NPE and NEE. We found for our applications that MAF worked better than more sophisticated architectures such as Neural Spline Flows \cite{durkan2019neural}. Like any density estimation technique, normalizing flows may struggle to scale up to very high dimensional data. They also assume the data are continuously distributed.

We outline marginal density estimation with normalizing flows since this process is the basis of NEE (Section~\ref{sec:nee}). Let $\bx$ be a random, continuous variable of dimension $d$ with an unknown probability distribution. Let $\bz$ denote a $d$-dimensional multivariate Gaussian. Suppose there exists an invertible, differentiable function $T$ such that $T(\bx) = \bz$. By the change of variables theorem, we can compute the density of $\bx$ as
\begin{equation}
    p(\bx) = p_{\bz}(T(\bx))  \left \lvert \det J_{T} \left ( T(\bx) \right )\right \rvert,
\end{equation}
where $p_{\bz}$ is the Gaussian density function and $J_{T}$ is the Jacobian matrix of partial derivatives of $T$ with respect to $\bx$. Similarly, we can generate samples from $p(\bx)$ by sampling Gaussian variables $\bz$ and running them through $T^{-1}$.

Let $p(\bx; \psi)$ denote a flow-based model of $\bx$, with $\psi$ denoting the parameters governing the flow, and let $p^*(\bx)$ denote the true distribution of $\bx$. As a function of $\psi$, the forward Kullback-Liebler (KL) divergence between $p^*(\bx)$ and $p(\bx; \psi)$ is
\begin{align}
    D_{KL}[p^*(\bx) \parallel p(\bx; \psi)] \\
= - \mathbf{E}_{p^*(\bx)}[\log p(\bx; \psi)] + C,
\end{align}
where $C$ is constant. While we have assumed $p^*(\bx)$ to be unknown, we can approximate the divergence if we have access to enough samples of $\bx$. From this, we derive the negative log-likelihood objective function in Equation~\ref{eq:nee-emp}.

\subsection{Neural Evidence Estimation}
\label{sec:nee}

% consider putting a visualization in here. NPE and NEE side by side

As the basis for simulation model criticism, we directly estimate a given simulator's evidence $p(x)$ from simulated data using normalizing flows. When training NPE or related SBI methods, we simulate many data-parameter pairs $(\bx_s, \bth_s)$ from the prior and simulator. The set of training simulations, $S_{\bx} = \{\bx_1, \ldots, \bx_S\}$, can be thought of as a sample drawn from the unknown marginal likelihood distribution $p(\bx).$ Therefore, we approximate $p(\bx)$ by training a marginal density estimator $h_\bx(\bx; \psi)$ on $S_\bx$, with $\psi$ denoting the learnable parameters of a normalizing flow. We describe the procedure in Algorithm~\ref{alg:nee}.

After fitting a marginal density estimator to simulated data, we use the Classifier Two-Sample Test (C2ST) to evaluate the flow's convergence \cite{lopez2016revisiting}. We construct a dataset that mixes ``authentic'' simulated data and ``synthetic'' data generated by the trained flow estimator and train a random forest to classify the data by origin. If the classifier's accuracy on held-out validation data is near random, then we deem our marginal likelihood approximation to be an adequate one. If, however, the classifier achieves better-than-random accuracy, then it is likely that the flow has not converged, and we should treat our marginal likelihood estimates with caution.

\begin{algorithm}
    \caption{Neural Evidence Estimation with Normalizing Flows}\label{alg:nee}
    \begin{algorithmic}
    \Require Prior distribution $p(\bth),$ simulator $p(\bx \mid \bth),$ marginal density flow estimator $h_{\bx}(\cdot \ ; \psi)$ observed data
    $\bx_o$ \Ensure Observed marginal likelihood estimate $\hat p(\bx_o)$
    \For{$s = 1, 2, \ldots S$} \State Sample $\bth_s \sim p(\bth)$ \State
    Simulate $\bx_s \sim p(\bx \mid \bth_s)$ \EndFor \State Using stochastic
    gradient descent, solve
    \begin{equation}\label{eq:nee-emp}
    \psi^* = \argmin_{\psi} \ - \frac{1}{S} \sum_{s=1}^S \log h_{\bx}(\bx_s; \psi)
    \end{equation}
    \State{$\hat p(\bx_o) \gets h_\bx(\bx_o; \psi^*)$}
    \end{algorithmic}
\end{algorithm}

Once NEE has been trained, it can produce a marginal likelihood estimate for any dataset with a single forward pass through the flow's network, which facilitates criticism using many observations.

\subsubsection{Hypothesis Testing for Diagnosing Model Misspecification}
\label{sec:hypo-testing}

Assuming we have arrived at a reliable estimate of the evidence, we can construct a statistical test to detect misspecification. For our null hypothesis, suppose that $\bx_o \sim p(\bx)$, i.e. that the observed data were generated by the simulation model for some parameter configuration. We compare the estimated marginal likelihood $\hat p(\bx_o)$ against a sampling distribution of estimated marginal likelihoods of many simulated datasets. For some large number $M$, simulate data points $\bx_1, \ldots, \bx_M$ from $p(\bx)$. We then compute a $p$-value:
\begin{equation}
    p = \frac{1}{M} \sum_{i=1}^M \mathbf{1}\{ p(\bx_i) < p(\bx_o) \}.
\end{equation}
\label{eq:compute-p}
Here, $p$ gives the (empirical) probability that a \textit{simulated} marginal likelihood will be smaller than the marginal likelihood for the observed data. Under the null hypothesis and for large $M$, $p$ has a roughly uniform distribution over the interval $(0, 1)$. 

We compute this $p$-value in practice by plugging in our normalizing flow estimator for $p(\bx)$ from Algorithm~\ref{alg:nee} and comparing it against the marginal likelihoods for the validation set of simulations. For some predetermined significance threshold, $\alpha \in (0, 1)$, if the computed $p$-value is less than $\alpha$, then we reject the null hypothesis and deem the proposed model to be misspecified. Informally, we judge the observed data to have low evidence under the stipulated model; $p$ should \textit{not} be interpreted as the probability that the model is misspecified. The threshold $\alpha$ gives the theoretical false positive rate across observations; if we are testing many models on one dataset, we recommend applying a suitable correction (e.g. Bonferroni) or else interpreting $\alpha$ with caution \cite{streiner2011correction, ranganathan2016common}.  

This test has previously appeared as an unpublished method in the \textbf{sbi} software library \cite{BoeltsDeistler_sbi_2025}. It is analogous to using Mahalanobis distance to detect outliers in multidimensional data \cite{etherington2019mahalanobis}. For multivariate normal distributions, the probability density of a point is uniquely determined by the Mahalanobis distance. If the observed data set is approximately multivariate normal, then the distribution of Mahalanobis distances will follow a $\chi^2$ distribution with the degrees of freedom equaling the dimensionality of the data. 

\subsubsection{Bayes Factors for Model Comparison}
\label{sec:bayes-factors}

Suppose we are faced with the problem of choosing among a discrete set of $K$ models. Bayesian model selection proceeds by treating the model itself as an unobserved, higher-order, parameter, or hyperparameter. Let $\mdl_k$ denote the model index, with $k \in \{1, \ldots, K \}$. For a given model $\mdl_k$, we can rewrite the posterior distribution of its unknown parameters (Equation~\ref{eq:posterior}) as follows:
\begin{equation}
    p(\bth \mid \bx_o, \mdl_k) = \frac{p(\bth \mid \mdl_k )\ p(\bx_o \mid \bth, \mdl_k)}{p(\bx_o \mid \mdl_k)}.
\end{equation}

Let us assign a discrete hyperprior $p(\mdl_k)$ over the space of candidate models. Through a second application of Bayes' theorem, we can see that
\begin{equation}
    p(\mdl_k \mid \bx_o) \propto p(\bx_o \mid \mdl_k)\ p(\mdl_k).
\end{equation}
Thus, the posterior probability of model $\mdl_k$ for observation $\bx_o$ is determined by the model's evidence $p(\bx_o \mid \mdl_k)$ and the hyperprior probability $p(\mdl_k)$. Suppose that all candidate models are equally probable \textit{a priori}, a common simplifying assumption. Then,
\begin{equation}
    p(\mdl_k \mid \bx_o) = \frac{p(\bx_o \mid \mdl_k)}{\sum_{l=1}^K (\bx_o \mid \mdl_l) }.
\end{equation}

The Bayesian approach toward uncertainty in model choice is to compute an average over the model posterior distribution \cite{fragoso2018bayesian}. For instance, the predictive distribution obtained from Bayesian model averaging is
\begin{equation}
    \label{eq:bma}
    p(\bx \mid \bx_o) = \sum_{k=1}^K p(\bx \mid \bx_o, \mdl_k) p(\mdl_k \mid \bx_o),
\end{equation}
where $p(\bx \mid \bx_o, \mdl_k)$ is the posterior predictive density for model $\mdl_k $ (see Equation~\ref{eq:ppd}). 

Model selection via Bayes factors can be seen as a computationally expedient alternative to Bayesian model averaging. Suppose we are interested in comparing two models, $\mdl_1$ and $\mdl_2$. AS before, we assign them equal prior probabilities. The ratio of these models' posterior probabilities is then
\begin{equation}
    B_{1,2} := \frac{p(\mdl_1 \mid \bx_o)}{p(\mdl_2 \mid \bx_o)} = \frac{p(\bx_o \mid \mdl_1)}{p(\bx_o \mid \mdl_2)}.
\end{equation}
This ratio between the two marginal likelihoods is called the \textit{Bayes factor}, and it measures the relative strength of evidence for $\mdl_1$ in comparison to $\mdl_2$ \cite{mackay1992bayesian}. For a qualitative interpretation of Bayes factors, we use the thresholds proposed by \textcite{jeffreys1998theory} in our experiments, shown in Table~\ref{tab:evidence-levels} (cf. \textcite{kass1995bayes}). 

\begin{table}
\centering
\begin{tabular}{@{}ll@{}}
\toprule
   $B$ & Evidence Level \\ \midrule
   $\leq 1$ & Negative \\
   $(1, 10^{1/2}]$& Negligible \\
   $(10^{1/2}, 10]$ & Substantial \\
   $(10, 10^{3/2}]$ & Strong \\
   $(10^{3/2}, 100]$ & Very Strong \\
  $>100$ & Decisive \\
\bottomrule 
\end{tabular}
\caption{Bayes factor ($B$) thresholds for evidence in favor of $\mdl_1$ vs $\mdl_2$, following \textcite{jeffreys1998theory}.}
\label{tab:evidence-levels}
\end{table}

\textcite{mackay1992bayesian} liken Bayes factors to a quantitative \textit{Ockham's Razor}, referring to the venerable logical principle that one should prefer a simpler explanation of a phenomenon to a more complex one if both are viable. In certain simple scenarios, it can be shown that Bayes factors automatically penalize model complexity, analogously to information criteria. Intuitively, the evidence corresponding to a simple model (i.e. one with few parameters) will have a more concentrated marginal likelihood, while a complex model (one with many parameters) will have a diffuse marginal likelihood density. The former strongly predicts a narrow set of realization, whereas the latter weakly predicts many outcomes. If both are sufficient to explain an observation $\bx_o$, then $p(\bx_o)$ will be greater for the simpler model. For further information on the statistical theory and epistemology of Bayesian model selection, see \textcite{mattei2019parsimonious}.

For all but the simplest Bayesian models, computing the evidence is a hard problem, since it corresponds to the unknown normalization constant for the posterior. The basic approach is the harmonic mean estimator,
\begin{equation}
\hat p(\bx_o) = \frac{1}{N} \sum_{i=1}^N \frac{1}{p(\bx_o \mid \bth_i)}
\end{equation}
for posterior samples $\bth_i$ \cite{newton1994approximate}. Though unbiased, this estimator is notorious for having an explosive variance since it is a sum of reciprocals of the likelihood density, and subsequent estimators have aimed to reduce this variance through strategies such as importance sampling, tempering, and Laplace approximation. For a review of traditional, sampling-based techniques for marginal likelihood estimation, see \textcite{llorente2023marginal}. For our purposes, we can approximate the Bayes factors for any pair of models without evaluating their likelihoods by plugging in the NEE estimates of evidence computed via Algorithm~\ref{alg:nee}. 

% In contrast to NEE, which offers quick, amortized inference for the evidence of any dataset $\bx$, these methods require running an entire sampling procedure for every new dataset.

\subsubsection{Related Work: Model Criticism}
\label{sec:related}

``No statistical model can safely be assumed adequate. Perspicacious criticism employing diagnostic checks must therefore be applied," wrote statistician George Box, in a variation on his most famous aphorism \cite{box1980sampling}. Every model is a working hypothesis; it should balance flexibility and parsimony so as to explain observations without making superfluous assumptions, neither underfitting nor overfitting the data \cite{box1976science}. In this section, we review common methods used for model comparison in epidemiology (and related disciplines) as well as the existing tools for model criticism in the setting of simulation-based inference. We argue that NEE addresses key shortcomings of both methodologies. 

Within the life sciences, despite recognition of the importance of multi-model analyses, there is little consensus around how to perform model comparison \cite{hooten_guide_2015}. The traditional approach to critical evaluation of data models has centered on statistical, likelihood-based analysis of residuals or errors in a model's predictions \cite{box1980sampling}. The widely used Akaike Information Criterion (AIC, see Section~\ref{sec:data-analysis}) gives an asymptotic approximation of a model's predictive error \cite{cavanaugh2019akaike}. AIC and other information criteria are functions of likelihood that penalize models with many parameters; they can favor under- or over-parametrized models at smaller sample sizes. When model likelihoods are unknown, as is often the case with complex stochastic models, visual or numeric predictive checks of in-sample residuals are often employed as a heuristic for comparison, though these can result in facile criticism. A more principled version of this approach is cross-validation (CV) of 
errors on held-out data \cite{bates2024cross}. CV can help strike a balance between overfitting and underfitting and is ubiquitous in machine learning, where predictive accuracy is generally at a premium compared to uncertainty quantification. However, CV can be computationally demanding (requiring fitting a model many times) and works best with independently sampled data. For data with dependency structure, like time series simulated from stochastic compartmental models, specialized procedures must be employed judiciously \cite{liu2024using}.

In recognition of the probability that any one model of an incompletely observed system is flawed, ensemble modeling---i.e. aggregating the outputs of multiple models---is widely used in epidemiology to generate robust predictions \cite{cramer2022evaluation}. Ensemble modeling proceeds by averaging the outputs of several models, often through a simple arithmetic mean or quantile averaging \cite{shandross2026multi}. Ensemble models in public health often show improved forecasting performance compared to individual models, but they can be difficult to interpret. Bayesian model averaging is arguably a more principled way to weigh the contributions from several models, as compared to unweighted averages or performance-based weighing, but it compounds the computational challenges of single-model Bayesian inference (see Section~\ref{sec:bayes-factors}).

% principled way to specify weights?

Approximate Bayesian Computation (ABC) is a well established statistical methodology for calibrating and comparing data models that bypasses direct evaluation of the likelihood and so can be seen as a forerunner of SBI.
While first developed for problems in population genetics \cite{beaumont2002approximate}, ABC has long been used in infectious disease epidemiology \cite{toni_approximate_2009}. The methods constituting ABC are fundamentally based on rejection sampling of simulated data, so it can become quite inefficient as the dimensionality of the data increase. Summary statistics are used to compress the data into a lower dimensional space, yet ABC model selection is considered highly sensitive to an insufficient set of summary statistics \cite{robert2011lack, marin2018likelihood}. Another established technique in infectious disease epidemiology, nearer to classical Bayesian inference, is Reversible Jump MCMC (RJMCMC). This procedure involves recovering intractable model likelihoods through Gibbs sampling (a.k.a. data augmentation) and using MCMC to estimate posterior model probabilities \cite{knock_bayesian_2014}. The difficulty arises when the MCMC sampler must ``jump'' between parameter spaces of different dimensionality. While in principle capable of (nearly) exact Bayesian inference, this methodology demands the design of elaborate sampling and proposal schemes that must be tailored to specific models. 

SBI synthesizes the uncertainty quantification of Bayesian statistics with the computational speed of neural networks and has shown promise for calibrating complex transmission models. However, SBI inherits a vulnerability to unreliable out-of-distribution prediction from deep learning (see Section~\ref{sec:sbi}). \textcite{ward_robust_2022} formalized the problem of model misspecification in the context of SBI and argued for the importance of a model criticism stage of analysis separate from parameter estimation. (If the two are not decoupled, then failure to calibrate a given model can be conflated with the model itself being misspecified.) To improve the robustness of NPE, they assigned an error model to summarized, low-dimensional simulation data that explicitly accounts for the simulation-to-reality gap and estimate the error posterior distribution. Their procedure, which combines MCMC sampling with the flow-based estimation of the prior predictive distribution (i.e. marginal likelihood), involves two potential sources of approximation error. This approach assumes that the simulator's error distribution belongs to a known family, such as a spike-and-slab model, which may be hard to justify for outputs beyond handcrafted summary statistics.

In a later work, \textcite{schmitt2023detecting} proposed quantifying simulation model misspecification via the Maximum Mean Discrepancy (MMD) between a model's marginal likelihood and the (unknown) true data generating process. Their proposed technique for detecting model misspecification involves prepending a ``summary network'' (a module that compacts input data into a low-dimensional space) to the NPE inference network. Then, a sample-based estimator of MMD is used to identify discrepancies between simulations and observations through frequentist hypothesis testing. Unless many independent and complete datasets are available---an improbable scenario in many applications of SBI, e.g. modeling a pandemic at the population level---the sample-based MMD estimator is noisy. They jointly train the summary and inference networks, coupling model criticism to parameter estimation, through a quasi-adversarial loss function based more in intuition than logic. 

Finally, \textcite{spuriomancini_bayesian_2023} introduced the problem of statistical model selection into SBI and developed a likelihood-free technique for estimating Bayes factors
using the targeted harmonic mean estimator for marginal likelihood. Their method entails estimating a surrogate likelihood density via Neural Likelihood Estimation (NLE) and a surrogate posterior (either via NPE or MCMC with the NLE surrogate), of which the latter serves as the target distribution for the harmonic mean estimator. Here again successful model criticism is contingent on correctly estimating the parameters.

Our method, NEE, distinguishes itself from previous approaches in several ways. NEE circumvents likelihood evaluation, allowing it to scale efficiently to large models and complex datasets. NEE offers a theoretically grounded, systematic solution to simulation model criticism, integrating the problems of diagnosing misspecification and model comparison. Like NPE, NEE performs amortized inference, so after an up-front training cost it can estimate the marginal likelihood of multiple datasets with ease. NEE avoids the need for either an an analytic likelihood density or a good approximation thereof. Since it makes few assumptions about the data it learns from, NEE is interoperable with a wide range of simulators. Compared to other deep learning-based approaches, NEE is a single-stage estimation procedure, making it simpler both conceptually and practically. By directly estimating marginal likelihood from a simulator's prior predictive distribution, NEE cuts a metaphorical Gordian knot, circumventing the need for high variance Monte Carlo estimators, involved sampling procedures, and surrogate likelihoods or posteriors. NEE decouples model criticism from parameter estimation, which mitigates the risk of confounding inference failures resulting from misspecification. 

% (In contrast, modern SBI techniques such as NPE leverage supervised machine learning to efficiently interpolate between simulations in high-dimensional space.)

\subsection{Stochastic Compartmental Models}
\label{sec:stochastic}

Stochastic compartmental models classically represent the transmission of an infectious disease through a population that is partitioned into distinct disease states, or compartments \cite{greenwood2009stochastic, britton_stochastic_2010, andersson2012stochastic}. The mechanisms that govern movement from one compartment to another model transitions as probabilistic events that depend on a set of parameters, which usually must be inferred from observed data. 
% maybe one sentence about how these exemplify some of the general challenge of complex simulation models. latency, nonlinear interactions

The well-known SIR model serves as a template for the transmission models considered in this work. Under the basic SIR model, a closed population of size $N$ is divided into three groups, susceptible ($S$), infected ($I$), and removed ($R$). We assume homogeneous, random mixing (contact) within the population. The SIR model was first conceived as a \textit{deterministic} system of ordinary differential equations \cite{kermack1927contribution}, and this remains its most widely implemented form. For time $t > 0$, we can model the changing sizes of the three compartments as follows:
\begin{align}
    & \frac{dS}{dt} = - \beta S\frac{I}{N} \\
    & \frac{dI}{dt} = \beta S\frac{I}{N} - \gamma I \\
    & \frac{dR}{dt} = \gamma I,
\end{align}
where $\beta$ is the infection rate and $\gamma$ is the recovery rate. Alternatively, $1/\gamma$ defines the average period of infectiousness.

The deterministic SIR model is a useful and accurate description of transmission dynamics in a large population. In smaller populations, we expect \textit{demographic uncertainty} to matter, motivating stochastic variants of the SIR model that explicitly represent the randomness of transmission.
% assumes homogeneous, random mixing within a population

\subsubsection{Continuous-Time Stochastic SIR Model}

Let $X(t)$ and $Y(t)$ be the number of individuals who have been infected and have recovered by time $t,$ respectively. These processes are sufficient to describe the state space of the SIR model, as $S(t) = N - X(t)$, $I(t) = X(t) - Y(t),$ and $R(t) = Y(t)$. We define the continuous-time stochastic SIR model to be a bivariate Markov jump process with two valid transitions. We first define the force of infection at time $t$ to be $\lambda(t) = \beta \frac{I(t)}{N}$. Let $\mathcal{H}_t = X(t), Y(t)$ denote the state of the SIR system at time $t$. For $h>0,$
\begin{align}
    P\left (X(t+h) - X(t) = 1 \mid \mathcal{H}_t \right) = \lambda(t)S(t)h + o(h) \\
    P(Y(t + h) - Y(t) = 1 \mid \mathcal{H}_t ) = \gamma I(t)h + o(h).
\end{align}
These equations define a pair of inhomogeneous Poisson processes in which the instantaneous probabilities of increase are approximately linear in $h$. Conversely, the waiting times between successive infections and recoveries are exponentially distributed.

\subsubsection{Discrete-Time Stochastic SIR Model}

Observations in an epidemic are often counts (e.g. of new infections or hospitalizations) reported at regular intervals such as days or weeks. This motivates a discrete-time formulation of the stochastic SIR model. For time steps $t=1, \ldots, T$, we define the following transition probabilities:
\begin{align}
    P(X(t+1) - X(t) = m \mid \mathcal{H}_t) =  \binom{S(t)}{m}(1 - e^{{-\lambda(t)}})^m(e^{-\lambda(t)})^{S(t) - m} \\
    P(Y(t+1) - Y(t) = n \mid \mathcal{H}_t) = \binom{I(t)}{n}(1 - e^{{-\gamma}})^n(e^{-\gamma})^{I(t) - n}.
\end{align}
The discrete-time stochastic SIR model is thus made up of a pair of binomial Markov chains. Transition probabilities are assumed to be constant between time intervals, so each Bernoulli trial (i.e. individual-level transition) can be thought of as a single exponential event over the unit interval.

\subsubsection{Simulation}

Let $\bX = \{X_t\}$ and $\bY = \{Y_t\}$ denote the complete processes for new infections and new recoveries, and let $\bth = (\beta, \gamma)$. If $\bX$ and $\bY$ are completely observed, then either version of the stochastic SIR model has a tractable, closed-form likelihood. However, epidemics are rarely completely observed in practice \cite{bu2022likelihood, morsomme2025exact}, motivating simulated-based inference among other methodologies for calibrating models with large latent spaces. 

For the sake of argument, suppose that only recovery times $\bY$ are observed.  To compute the posterior $p(\bth \mid \bY)$ through likelihood-based inference, we would need to solve for the partial likelihood $p(\bY \mid \bth)$. However, 
\begin{equation}
    p(\bY \mid \bth) = \int p(\bX, \bY \mid \bth) d\bX
\end{equation}
a high-dimensional integral over the complete likelihood. Deriving the partial likelihood involves marginalizing over all the possible trajectories of $\bX$, which can be exponentially complex given the dependency of $\lambda(t)$ on $\bX$ and $\bY$.

While solving the stochastic SIR model's partial likelihood analytically is infeasible at scale, sampling trajectories via computer simulation is relatively straightforward and inexpensive. The discrete-time variant of the SIR model has a more awkward expression but is easier to implement as a simulation. To sample from the continuous-time SIR model, Gillespie's algorithm \cite{gillespie1976general} is commonly used. When the time increments are small, relative to $\beta$ and $\gamma$, the discrete-time model will behave similarly to its continuous-time counterpart.

\section{Results}

\subsection{Simulation Experiment I: Selecting among SIR-like Compartmental Models}
\label{sec:experiment1}

Suppose we observe the incidence of some infectious disease over 10 discrete time steps within a closed population of size 100. Our data take the form $\bx = X_1, \ldots, X_{10}$, where each $X_t$ is the proportion of subjects who have shown symptoms at or before time step $t$. We assume complete observation of cases. We consider five different stochastic compartmental models:

\begin{itemize}
    \item $\mdl_0$: a Susceptible-Infected-Recovered (SIR) model;
    \item $\mdl_1$: a SIR model with \textit{constant hazards}, such that the force of infection remains constant, independent of the frequency of infected patients $I_t/N$;
    \item $\mdl_2$: a Susceptible-Infected (SI) model in which infected patients remain infectious indefinitely;
    \item $\mdl_3$: a susceptible-infected-recovered-susceptible (SIRS) model, in which recovered patients randomly regain susceptibility. In this model, the proportion of infected patients can exceed 1, to account for reinfections
    \item $\mdl_4$: a susceptible-exposed-infected-recovered (SEIR) model, which adds a latent, incubation state where individuals have been exposed to infection but are not yet infectious themselves.
\end{itemize}

\begin{figure}
\begin{subfigure}[t]{.45\textwidth}
    \centering
    \includegraphics[width=\linewidth]{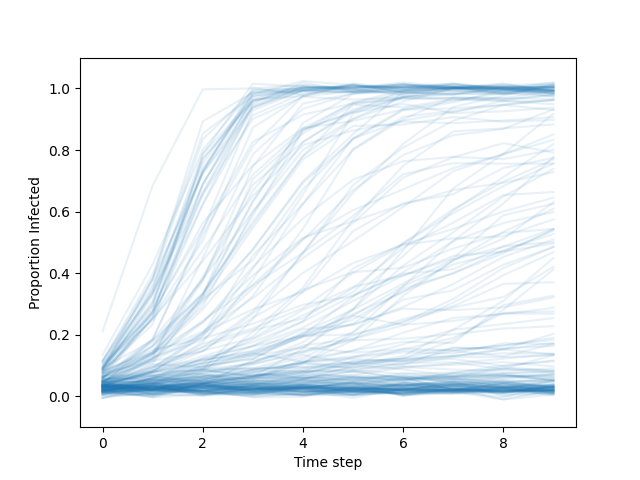}
    \caption{SIR Model}
\end{subfigure}
    \hfill
\begin{subfigure}[t]{.45\textwidth}
    \centering
    \includegraphics[width=\linewidth]{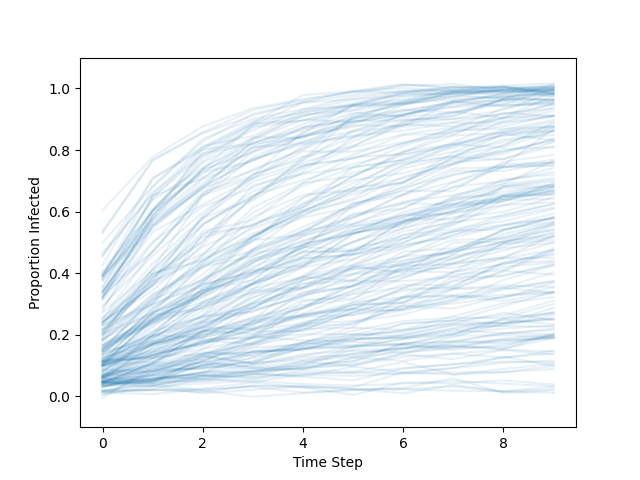}
    \caption{Constant Hazards SIR Model}
\end{subfigure}
\medskip
\begin{subfigure}[t]{.45\textwidth}
    \centering
    \includegraphics[width=\linewidth]{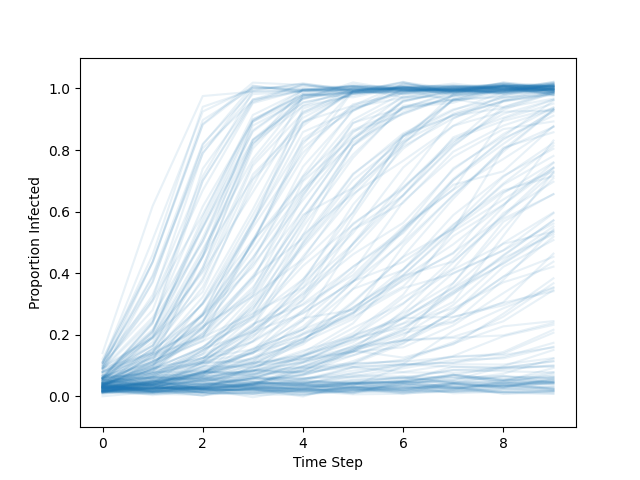}
    \caption{SI Model}
\end{subfigure}
    \hfill
\begin{subfigure}[t]{.45\textwidth}
    \centering
    \includegraphics[width=\linewidth]{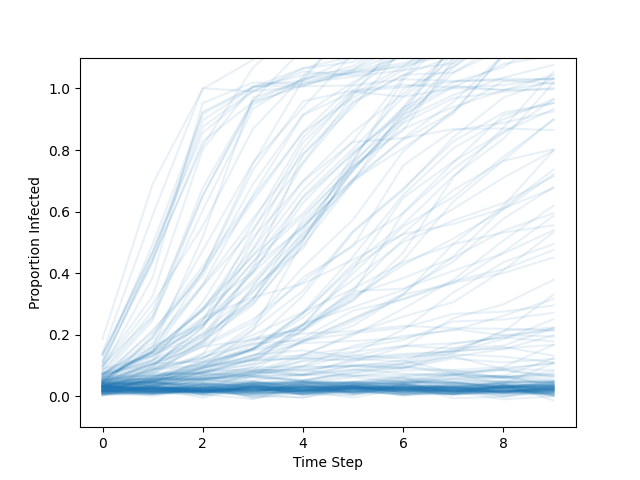}
    \caption{SIRS Model}
\end{subfigure}
\medskip
\begin{subfigure}[t]{.45\textwidth}
    \centering
    \includegraphics[width=\linewidth]{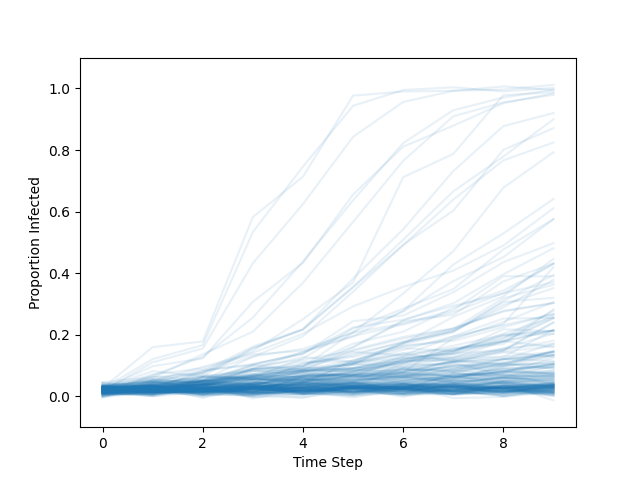}
    \caption{SEIR Model}
\end{subfigure}
\caption{Visualization of the marginal likelihood distribution for five stochastic epidemic models. Each plot shows 200 realizations of the proportion of infected individuals over time.}
\label{fig:marginals}
\end{figure}

Figure~\ref{fig:marginals} visualizes the marginal likelihood distributions for these five models by plotting 200 simulation draws from each. Information on the models' parameters and the priors we used can be found in the supplementary materials. This experiment, which considers the problem of comparing multiple compartmental models, is comparable to one found in \textcite{toni_approximate_2009}. Though clinical data might suffice to determine basic properties of an epidemic---e.g. is infectiousness temporary or permanent?---we can say that this simulation experiment suggests the very early stages of an emergent epidemic in which the fundamental dynamics are still unknown. 

For each model, we trained a flow-based marginal density estimator on 2,000 simulations of $\bx$. We added a small amount of independent Gaussian noise $\epsilon_t \sim \mathcal{N}(0, 0.1)$ to each point $X_t$ to ensure the flows' convergence. We took the SIR model $\mdl_0$ to be the true model and simulated 1,000 ``observed'' datasets. In each of the 1,000 trials, we estimated the marginal likelihood of $p(\bx_o ; \mdl_i)$ under each model $\mdl_i$ and computed the $p$-value for the null hypothesis that $\bx_o \sim p(\bx ; \mdl_i)$. Since we are running five tests in each trial, we apply the Benjamini-Hochberg adjustment to $p$-values to control the false positive (i.e. rejection) rate.

Table~\ref{tab:hyp-testing-sim} reports the rejection rate for each candidate model computed over 1,000 trials at three different significance thresholds, $\alpha=0.1, 0.05, 0.025$. At all levels of $\alpha$, this hypothesis testing procedure correctly flags the constant hazards and SI models as misspecified at least 40\% of the time, with this fraction increasing for larger values of $\alpha$. The rejection rate is substantial, but nonetheless lower, for the SEIR model, particularly for $\alpha=0.05$ and smaller. The rejection rate for the SIRS model is similar to $\alpha$, suggesting that the test treats the SIRS model as essentially correctly specified. We hypothesize that when the rate at which individuals lose immunity is very small, a realization of the SIRS model is indistinguishable from an SIR draw. Similarly, when the incubation period is very short, an SEIR trajectory looks similar to an SIR trajectory. We could conceive of the SIR model as a special case of the SIRS and SEIR models. 

Overall, this test is somewhat underpowered with respect to these datasets: in the majority of all trials, we fail to reject the incorrect models. For the correct model, we observe that the test's empirical false positive rate is slightly lower than the notional false positive rate ($\alpha$) at significance levels $\alpha=0.1$ and $0.05$. However, we judged this to be the result of sampling error resulting from the finite set ($M = 400$, see Equation~\ref{eq:compute-p}) we use to estimate $p$-values: on average, the false positive rate is well-calibrated for varying levels of $\alpha$ (see supplemental materials).

\begin{table}
\centering
\begin{tabular}{@{}llll@{}}
\toprule
Model            & $\alpha=0.1$ & $\alpha=0.05$ & $\alpha=0.025$ \\ \midrule
\textbf{SIR}     & 0.065        & 0.036         & 0.021          \\
Constant Hazards & 0.426        & 0.420         & 0.414          \\
SI               & 0.424        & 0.419         & 0.414          \\
SIRS             & 0.088        & 0.037         & 0.012          \\
SEIR             & 0.394        & 0.351         & 0.324          \\ \bottomrule
\end{tabular}
\caption{Rejection rates for the model misspecification test at different significance levels. Results shown for 1,000 trials. We use the Benjamini-Hochberg correction for multiple testing to adjust $p$-values.}
\label{tab:hyp-testing-sim}
\end{table}

\begin{figure}
    \centering
    \includegraphics[width=0.5\linewidth]{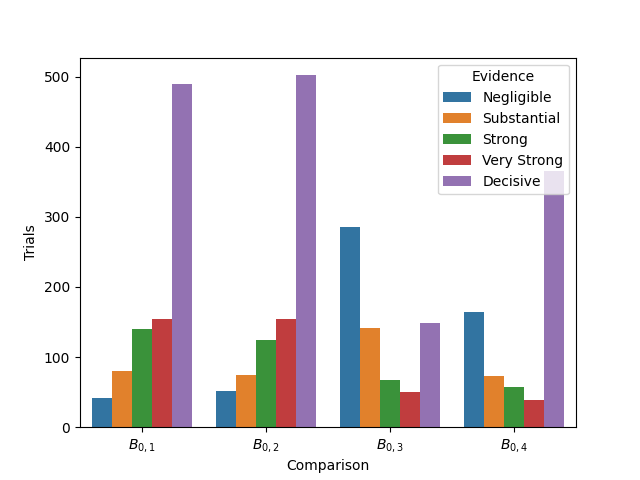}
    \caption{Bayes Factor comparisons between contending model pairs, where model 0 is the true model. We show evidence levels (see Table~\ref{tab:evidence-levels}) for 1,000 trials.}
    \label{fig:bayes-factors}
\end{figure}

As a complement to our hypothesis testing, we computed the Bayes factors for evidence in favor of the true model $\mdl_0$ relative to the false models $\mdl_1, \ldots, \mdl_4$ for each synthetic observation.  Then, using the thresholds listed in Table~\ref{tab:evidence-levels}, we tallied the evidence levels for each comparison. We show the resulting distribution in Figure~\ref{fig:bayes-factors}. We found decisive evidence in favor of the correct model over the constant hazards, SI, and SEIR model between 40\% and 50\% of the time, a rate comparable to the corresponding rejection rates found by the misspecification test. Likewise, the evidence in favor of the SIR model over the SIRS model is mainly negligible.

Finally, we train a neural posterior estimator on 2,000 simulation-parameter pairs for each of the five models. For 20 pseudo-observations generated from $\mdl_0$, we estimated the posterior under each model and ran a posterior predictive check. Table~\ref{tab:sir-msppe} shows the mean squared posterior predictive error (MSPPE) averaged across the 1,000 trials. We define the MSPPE to be
\begin{equation}
    \mathbf{E}_{p(\bx \mid \bx_o)}[(\bx - \bx_o)^2],
\end{equation}
an expectation that can be approximated from a large enough posterior predictive sample. Since our data are time series, we also average the MSPPE across the 10 time steps.

\begin{table}
    \centering
\begin{tabular}{lr}
\toprule
Model & MSPPE \\
\midrule
\textbf{SIR} & 0.163 \\
Constant Hazards & 0.0851 \\
SI & 0.128 \\
SIRS & 0.120 \\
SEIR & 0.912 \\
\bottomrule
\label{tab:msppe}
\end{tabular}
\caption{Mean squared posterior predictive error by model, averaged across 1,000 trials.}
    \label{tab:sir-msppe}
\end{table}

This table illustrates the pitfalls of performing model criticism through posterior predictive checks alone. The MSPPE for the true model, SIR, is the second highest, whereas the Constant Hazards SIR model, which misspecifies the basic mechanism of frequency-dependent transmission, has the lowest average error. The SEIR model exhibits by far the highest MSPPE. It appears that MSPPE as a selection criterion may favor simple, low-variance models like Constant Hazards and conversely penalize flexible, high-variance models such as SEIR. Had we performed model selection by picking the model with the lowest MSPPE, we would have chosen Constant Hazards in 90\% of trials and SI in the remaining 10\%.  

\subsection{Simulation Experiment II: Tuning Transmission Heterogeneity}
\label{sec:experiment2}

As before, suppose we observe the incidence of some epidemic over 10 time intervals  in a population of size 100. We extend the previous setup by supposing that we additionally observe individual-level demographic \textit{metadata}. Suppose we know the age of all our subjects along with the time steps at which they first show symptoms. This time, we assume that symptom onset times can be taken as a proxy for recovery times (a realistic assumption if, e.g., symptomatic individuals are placed in quarantine). A question that often arises in infectious disease modeling is whether transmission is heterogeneous or homogeneous \cite{knock_bayesian_2014}. In this instance, we are interested in determining whether susceptibility varies by age. We divide the population into five age cohorts and model the force of infection acting on individual $i$ at time t as
\begin{equation}
    \lambda_i(t) = \frac{\beta_{k(i)} I_t}{N},
    \label{eq:hetero-hazard}
\end{equation}
where $k(i) \in \{1, 2, 3, 4, 5\}$ is the age cohort of subject $i$. This formulation results in a heterogeneous SIR model with six parameters: the recovery rate $\gamma$ and five infection rates $\beta_1, \ldots, \beta_5 =: \bbeta$. When the rates are equal, this reduces back to the homogeneous model SIR model seen in the previous section.

We might be tempted to take the incidence time series within each age bin as our observation, i.e. $X^{(k)}_1, \ldots, X_{10}^{(k)}$ for $i$ for $k = 1, \ldots, 5$. This would result in data of dimension 50. However, estimating the density of such high dimensional data is a challenging task. Instead, we treat the \textit{size} of the epidemic, segmented by age, as our data. That is, $\bx = X^{(1)}_{10}, \ldots, X^{(5)}_{10}$, a more manageable vector of length five. We interpret this as the total number of recovered individuals in each age cohort. This \textit{final outcome} analysis approach is commonly used in epidemiology to model heterogeneous mixing, e.g. at the household level, or when accurate reporting times are unavailable \cite{knock_bayesian_2014, kypraios2017tutorial}.

\begin{figure}
    \centering
    \includegraphics[width=0.5\linewidth]{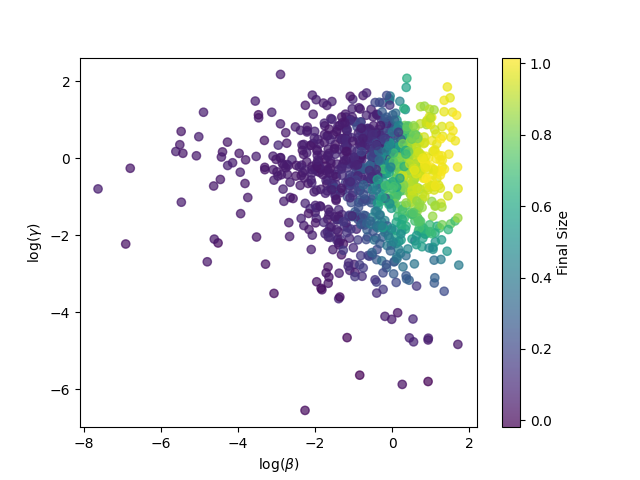}
    \caption{1,000 simulations of an epidemic's final size (as a proportion of the population) under a homogeneous SIR model along with input parameters.}
    \label{fig:final-size}
\end{figure}

We consider two simulation models: Model A, a SIR model with homogeneous mixing, and Model B, a SIR model with heterogeneous mixing by age group (Equation~\ref{eq:hetero-hazard}). Figure~\ref{fig:final-size} illustrates the relationship between the input parameters $\beta$ (transmission rate) and $\gamma$ (recovery rate) and the final size of the epidemic, aggregated across age brackets, for the homogeneous Model A. Higher values of $\beta$ clearly correlate with a greater number of total cases, but the total number of cases is less informative of $\gamma$. Since we do not observe the evolution of the epidemic over time in this experiment, recovery times cannot be readily inferred from the final outcome data.

Similarly to Section~\ref{sec:experiment1}, we simulate 1000 ``observed'' datasets from models A and B. We train normalizing flows on 2,000 simulations from each model to learn neural posterior and marginal likelihood estimators. We then apply models A and B to each of the 2,000 pseudo-observations, computing the MSPPE, the $p$-value associated with the misspecification test, and the Bayes Factor for the binary comparison between models A and B. Since we are running two tests per trial, we employ the conservative yet simple Bonferroni adjustment to $p$-values.

Table~\ref{tab:binary-comparison} reports the results of each of these model selection criteria averaged across 1,000 trials for each model. The first set of columns shows MSPPE, which measures the posterior predictive accuracy of a fitted model with respect to observed data. When the underlying data is heterogeneous (sampled from Model B), the MSPPE is about twice as high for Model A as for Model B. Model A lacks the flexibility to handle heterogeneous outcomes, so its predictions will often be biased, increasing MSPPE. However, when the data is homogeneous, the MSPPE is only slightly lower for Model A than for Model B. If we correctly fit model B to homogeneous data, the five transmission rate parameters should be close to one another, thus approximating the homogeneous model. Model B's predictions will still have higher variance, which may explain why its MSPPE is slightly larger. If we were to naively rely on MSPPE for model selection, we'd often default to selecting the more complicated heterogeneous transmission model even for homogeneous data.

The second set of columns shows the proportion of rejections of the null hypothesis of correct model specification at a significance threshold of $\alpha=0.05$. When the underlying data are homogeneous, we reject the heterogeneous model only 3\% of the time, while almost never rejecting the homogeneous model. The rejection rate for Model A is 0.014, close to but somewhat less than the Bonferroni-corrected threshold of $\alpha / 2 = 0.025$. Since Model A can be thought of as a special case of Model B, it is unsurprising that Model B is rarely outright misspecified for homogeneous data. Conversely, when the data results from heterogeneous mixing, we reject the homogeneous model about half the time, since it lacks the capacity to generate different outcomes by age group. 

In the third and final set of columns, we list the selection rate using Bayes Factors for a binary comparison between Models A and B. If the Bayes factor $B$ for one model versus the other is greater than $10^{0.5} \approx 3.2$ (a \textit{substantial} degree of evidence according to Table~\ref{tab:evidence-levels}), we select that model as the better fit for the observed data. For both homogeneous and heterogeneous outcome data, this Bayes factor procedure selects the correct model more than half the time, and selects the incorrect model between 6-10\% of the time. (In a minority of cases, the Bayes factor offers no evidence in favor of either model.) This demonstrates that Bayes factors not only help rule out misspecified models but encourage parsimony: though Models A and B may predict homogeneous data comparably well, the Bayes factor points towards the model with fewer parameters.  

\begin{table}
\centering
\begin{tabular}{@{}lllllll@{}}
\toprule
True Model    & \multicolumn{2}{l}{MSPPE} & \multicolumn{2}{l}{Rejection Rate} & \multicolumn{2}{l}{Selection Rate} \\ \midrule
              & A         & B       & A             & B            & A             & B            \\
A   & 0.0997       & 0.121      & 0.014             & 0.03            & 0.532             & 0.061            \\
B & 0.273        & 0.141      & 0.495             & 0.026            & 0.103             & 0.648            \\ \bottomrule
\end{tabular}
\caption{Model criticism metrics for a binary decision between a homogeneous transmission model (A) and a heterogeneous model (B) reported over 1,000 trials. Rejection Rate gives the proportion of $p$-values from the model misspecification test rejected at a significance level of $0.05$ (with a simple Bonferroni correction); Selection Rate reports the proportion of Bayes factors greater than 3.2 (i.e. a substantial degree of evidence per Table~\ref{tab:evidence-levels}).}
\label{tab:binary-comparison}
\end{table}

\subsection{Case Study: Explaining Reinfection in an Isolated Influenza Outbreak}
\label{sec:data-analysis}

Epidemics of respiratory viruses, such as influenza and the coronavirus, commonly occur in multiple waves. Over the course of the COVID-19 pandemic, the virus evolved into novel variants, against which prior infections provided limited immunity, resulting in widespread reinfections. For seasonal influenza, in contrast, the mechanisms producing multiple waves are less well understood, complicating the planning of public health responses to outbreaks \cite{mummert2013perspective}. One common hypothesis regarding multiple-wave epidemics posits the co-circulation of multiple, antigenically distinct strains that confer only partial cross-immunity \cite{andreasen1997dynamics, casagrandi2006sirc, barry2008cross, rios2009qualitative}. For past epidemics, virological and serological samples are scarce or non-existent, making it hard to directly verify this theory. An alternative hypothesis considers the heterogeneity of hosts---rather than pathogens---as the primary driver of reinfection: infection by influenza is theorized to provide hosts with varying degrees of short- or long-term immunity \cite{gomes2004infection, mathews2007biological, mathews2010prior, miller2010incidence}. Transmission models built around either of these theories may offer a reasonable fit to a given epidemic dataset, raising the question of what theory to support.

In this case study, we use NEE to perform model comparison for a multiwave 1971 influenza epidemic on the island of Tristan da Cunha, thereby comparing several distinct immunological mechanisms. We build upon a pioneering work by \textcite{camacho_explaining_2011} that used maximum likelihood analysis to evaluate six distinct transmission models offering competing theories of influenza reinfection. In the following section, we describe their dataset, the models they proposed, and their methodology, which we adapt to neural SBI.

\subsubsection{Dataset and Modeling}

\begin{figure}
    \centering
    \includegraphics[width=0.5\linewidth]{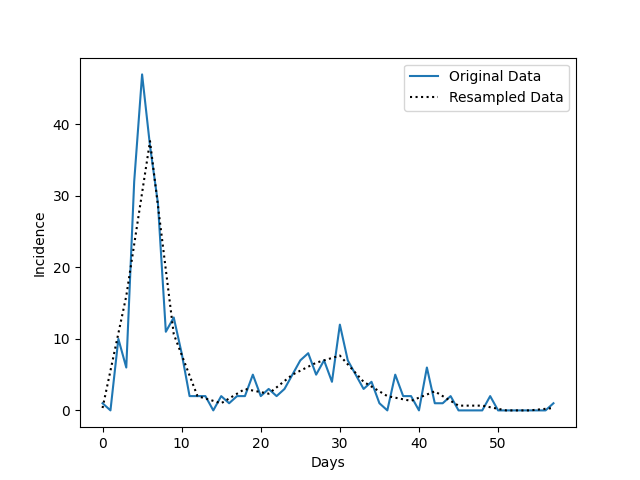}
    \caption{Observed incidence during the 1971 Tristan da Cunha influenza outbreak. We show both the original data---a daily time series---and the resampled data, a rolling average of incidence across three-day windows.}
    \label{fig:tdc-incidence}
\end{figure}

Tristan da Cunha (TdC) is a volcanic island and British Overseas Territory in the South Atlantic. Thanks to its remote location, contacts with the outside world are infrequent. On August 13th, 1971, a ship arrived at TdC, and five islanders disembarked. These islanders showed respiratory disease symptoms either during the voyage or immediately after. Soon an epidemic spread across the island, lasting 59 days (Figure~\ref{fig:tdc-incidence}). Incidence sharply peaked within the first week, but a second wave of infections occurred three weeks into the outbreak. (White multiple waves are a common phenomenon in influenza pandemics, this dataset is somewhat unusual in that the interval between the waves lasted days, not months.) Out of a total population of 284, 273 islanders were confirmed to have been infected, and 92 (about 1 in 3) experienced two attacks. The exact day of symptom onset are available for 312 out of 365 of the cases, making up our dataset. Later serological analysis of islanders revealed elevated antibodies against A/H3N2, a recently-evolved influenza subtype, though no virological study was conducted linking recrudescence to antigenically distinct strains.

The 1971 TdC influenza epidemic is a valuable case study for infectious disease modeling, since it features an remarkably closed population, relatively complete incidence reporting, and a two-wave structure with no obvious scientific explanation. \textcite{camacho_explaining_2011} propose six reinfection mechanisms, each corresponding to a modification of the well-known Susceptible-Exposed-Infected-Recovered (SEIR) compartmental model. (Infection by influenza results in a latent period, usually lasting several days, during which the host is asymptomatic and not yet infectious. For this reason, the SEIR model is more suitable than the SIR model for analyzing the 1971 TdC epidemic.)

We summarize the six immunological hypotheses as follows:
\begin{itemize}
    \item \textit{2 Virus} (\textbf{2Vi}): two separate viruses with differing transmissability were introduced into TdC at the beginning of the outbreak.
    \item \textit{Mutation} (\textbf{Mut}): during the first wave, a specimen of the virus mutated into a new antigenic variant and spread.
    \item \textit{All-or-Nothing} (\textbf{AoN}): following recovery from infection, some hosts acquired complete immunity, but others remained susceptible to infection by the same strain.
    \item \textit{Partially Protective Immunity} (\textbf{PPI}): recovered hosts develop a limited immunity that reduces the risk of reinfection.
    \item \textit{In-Host} (\textbf{InH}): some hosts are unable to completely eliminate the viral load post-infection and experience a second infectious period, independent of external contact.
    \item \textit{Window-of-Reinfection} (\textbf{Win}): following recovery, hosts take some time to acquire long-term immunity, making them susceptible to reinfection within a certain window. 
\end{itemize}
In addition to these reinfection models, we consider a seventh mechanism as a naive baseline:
\begin{itemize}
    \item \textit{SEIR Baseline} (\textbf{Base}): Recovery from infection grants immediate and complete immunity, so reinfections are impossible.
\end{itemize}

The data record the daily incidence, so we have information on the number of infectious individuals over time. All other model states are unobserved, greatly complicating likelihood-based inference. We follow \textcite{camacho_explaining_2011} in implementing these hypotheses as stochastic, continuous-time Markov chain simulators. We would expect demographic uncertainty to influence the outbreak's dynamics in light of TdC's minute population size. As in the original study, we sample continuous transition times using Gillespie's algorithm, and we adopt a Poisson process observation model to account for incomplete reporting and possible asymptomatic cases. The output of our simulation is a time series $\by = y_1, \ldots, y_{59}$, representing the observed daily incidence, scaled to the unit interval $[0, 1]$. For each $t=1, \ldots, 59$,
\begin{equation}
    y_t \sim \frac{1}{N} \text{Poisson}(\rho I_t),
\end{equation}\label{eq:obs_model}
where $N = 284$ is the total population of TdC, $I_t$ is the true number of infected at time $t$, and $\rho \in (0, 1)$ is the average reporting rate. Like \textcite{camacho_explaining_2011}, we treat $\rho$ as a parameter to be inferred from the data.

\textcite{camacho_explaining_2011} use Iterated Filtering \cite{ionides2006inference}, a sequential Monte Carlo procedure, to approximate the maximum likelihood estimate of unobserved parameters, including transition rates and the initial conditions (e.g. the number of infecteds at the start of the observation period). To perform model selection, they plugged the estimated maximum log-likelihood into the corrected Akaike information criterion (AIC), an estimator of prediction error that favors models with a greater likelihood and fewer parameters. We use their estimates of the initial conditions for each model but infer transition rates and the reporting rate from the data. We provide a description of all model parameters and our choice of prior distributions in the supplementary materials.

% mention: the original study, (Mantle & Tyrell 1973), dismissed the two-virus theory...

\subsubsection{Simulation-based Inference and Criticism}

\begin{figure}
\begin{subfigure}[t]{.30\textwidth}
    \centering
    \includegraphics[width=\linewidth]{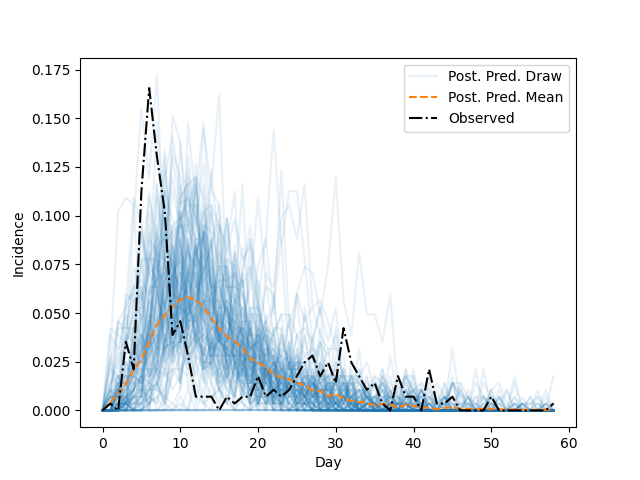}
    \caption{Base SEIR model}
\end{subfigure}
    \hfill
\begin{subfigure}[t]{.30\textwidth}
    \centering
    \includegraphics[width=\linewidth]{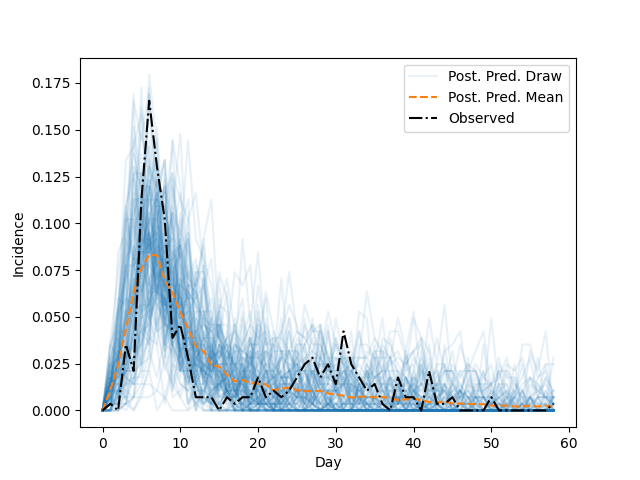}
    \caption{2Vi model}
\end{subfigure}
    \hfill
\begin{subfigure}[t]{.30\textwidth}
    \centering
    \includegraphics[width=\linewidth]{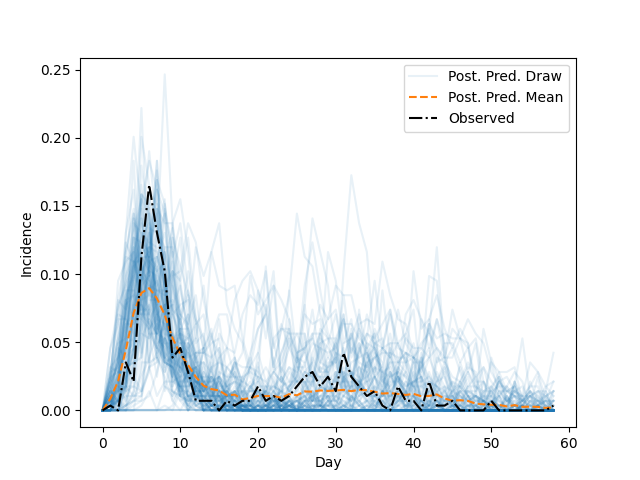}
    \caption{Mut model}
\end{subfigure}

\medskip

\begin{subfigure}[t]{.30\textwidth}
    \centering
    \includegraphics[width=\linewidth]{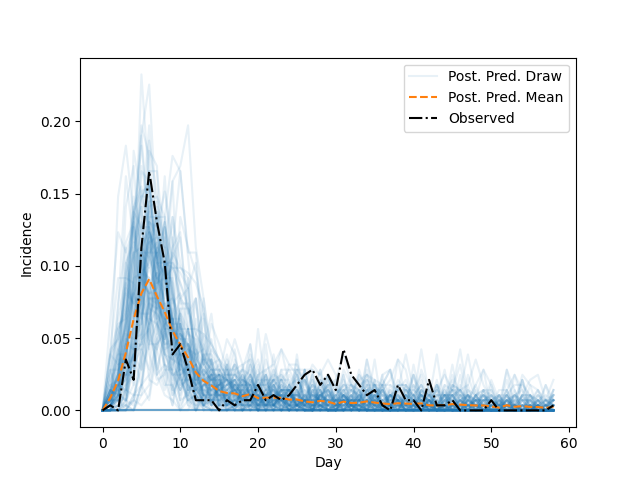}
    \caption{AoN model}
\end{subfigure}
    \hfill
\begin{subfigure}[t]{.30\textwidth}
    \centering
    \includegraphics[width=\linewidth]{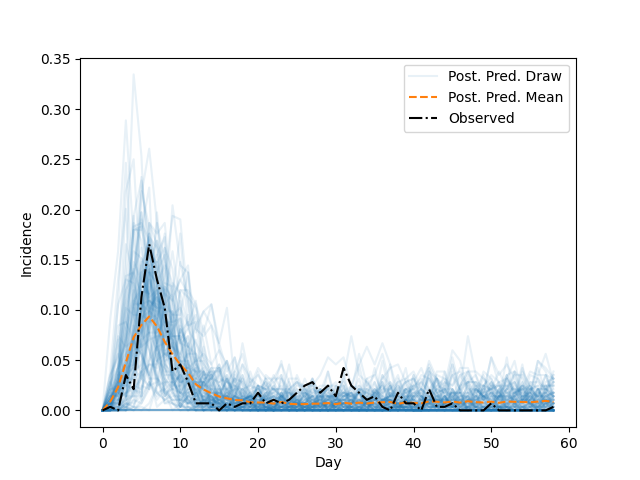}
    \caption{PPI model}
\end{subfigure}
    \hfill
\begin{subfigure}[t]{.30\textwidth}
    \centering
    \includegraphics[width=\linewidth]{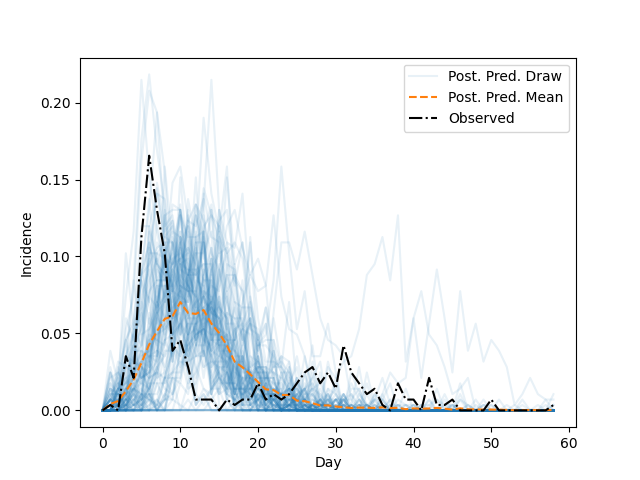}
    \caption{InH model}
\end{subfigure}

\medskip 

% \begin{subfigure}[t]{.30\textwidth}
%     \centering
%     \includegraphics[width=\linewidth]{figures/posterior_predictive_check_win.png}
%     \caption{Win model}
% \end{subfigure}

% \hfill

% \begin{subfigure}[t]{.30\textwidth}
%     \centering
%     \includegraphics[width=\linewidth]{figures/posterior_predictive_check_ensemble.png}
%     \caption{Ensemble}
% \end{subfigure}

\begin{subfigure}[t]{.30\textwidth}
    \centering
    \includegraphics[width=\linewidth]{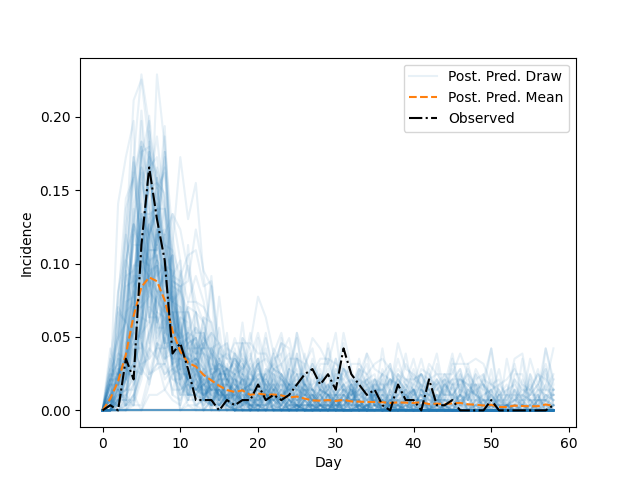}
    \caption{Win model}
\end{subfigure}
    \hfill
% \begin{subfigure}[t]{.30\textwidth}
%     \centering
%     \includegraphics[width=\linewidth]{figures/posterior_predictive_check_ppi.png}
%     \caption{PPI model}
% \end{subfigure}
    \hfill
\begin{subfigure}[t]{.30\textwidth}
    \centering
    \includegraphics[width=\linewidth]{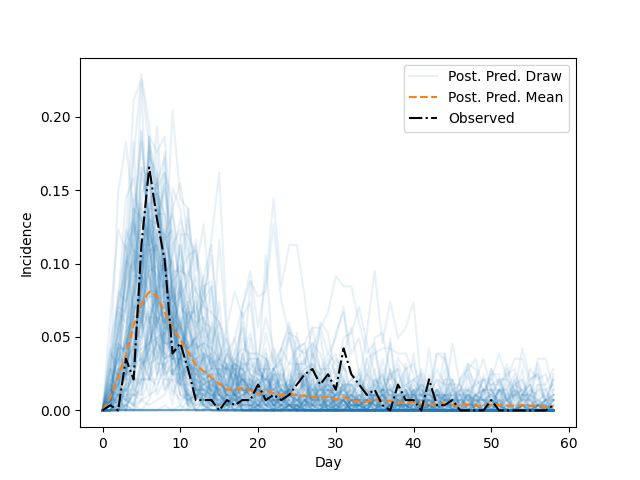}
    \caption{Ensemble}
\end{subfigure}

\caption{Posterior predictive checks of the fitted reinfection models.}
\label{fig:tdc-predictive-checks}
\end{figure}

For the first stage of data analysis, we fit the seven models to the observed data using NPE. We trained the neural posterior estimators on 2,000 simulation-parameter pairs drawn from each model. Figure~\ref{fig:tdc-predictive-checks}(a-g) depicts posterior predictive checks for each fitted model. At a glance, the Base and InH models appear to provide a poor fit to the observed data, with predicted incidence peaks around day 12, several days after the factual first wave peak. Mut is unique among the seven models in predicting, on average, a (slight) resurgence in cases aligning with the factual second wave between days 20 and 40. Mut and 2Vi, the dual virus mechanisms, exhibit greater predictive volatility than AoN, PPI, and Win, which posit heterogeneity of hosts rather than of the virus.

% discussing residuals/errors
Next, we analyze the residuals between the posterior predictions and the observed data. For all seven models, the posterior predictions were virtually unbiased, aggregated across all 59 days. That is, 
\begin{equation}
    \frac{1}{T}\sum_{t=1}^{59}\mathbf{E}_{p(\by \mid \by_o)}[y_i - y_{o, i}] \approx 0
\end{equation}
for all models. This is indicative of the predictive accuracy and flexibility of NPE.

Table~\ref{tab:tdc-errors} shows the MSPPE for each model computed across three time periods: the entire observation period, the second wave and the extinction period. MSPPE provides a quantitative measure of both the bias and variance of the predictive residuals over time. Table~\ref{tab:tdc-errors} affirms that Base and InH fit the observations poorly. The other five models achieve similar total accuracy. PPI has poor accuracy during the second wave period, whereas 2Vi seems to have the best accuracy in each of the three periods under consideration. Nonetheless, no model stands out as a clear favorite based on predictive accuracy.

\begin{table}
    \centering
\begin{tabular}{lrrr}
\toprule
Model & Total Error & Second Wave Error & Extinction Period Error \\
\midrule
Base & 0.0336 & 0.0327 & 0.0359 \\
2Vi & 0.0244 & 0.0256 & 0.0260 \\
Mut & 0.0266 & 0.0268 & 0.0301 \\
AoN & 0.0244 & 0.0257 & 0.0305 \\
PPI & 0.0276 & 0.0324 & 0.0253 \\
InH & 0.0352 & 0.0357 & 0.0374 \\
Win & 0.0255 & 0.0267 & 0.0290 \\
Ensemble & 0.0273 & 0.0320 & 0.0286 \\
\bottomrule
\end{tabular}

    \caption{Mean squared posterior predictive error reported across three time windows: the total period (days 1-59), the second wave (days 21-36), and the extinction period (days 52-59)}
    \label{tab:tdc-errors}
\end{table}

\begin{figure}
    \centering
    \includegraphics[width=0.5\linewidth]{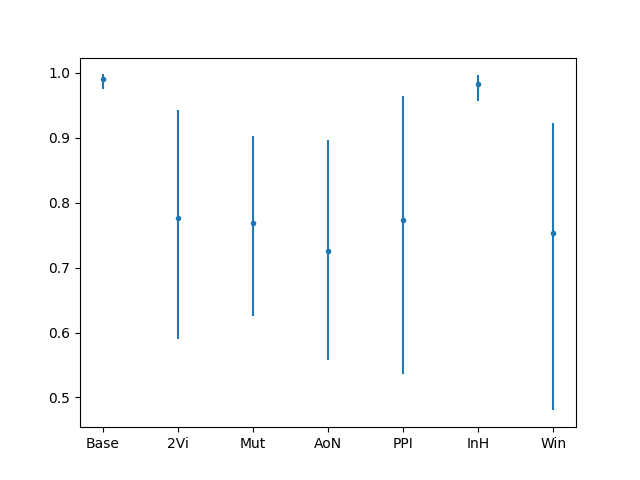}
    \caption{Posterior estimates of reporting rate $\rho$, showing the means and 90\% credible intervals.}
    \label{fig:rho_estimate}
\end{figure}

Figure~\ref{fig:rho_estimate} shows the posterior mean estimates of the reporting rate parameter $\rho$ for each model, along with 90\% credible intervals (bounded by the $(0.05, 0.95)$ quantiles). Except for InH and Base, where NPE inferred near-complete reporting with high confidence, our analysis detected major uncertainty as to the extent of incidence underreporting. The Poisson observation model (Equation~\ref{eq:obs_model}) assumed by \textcite{camacho_explaining_2011} is not unproblematic: in many cases it is challenging, if not impossible, to infer the incompleteness of data from the data themselves. A small observed incidence can be explained by either low infectiousness or underreporting, so $\rho$ and the other parameters may be jointly unidentifiable. We assumed a rather weak prior, $\rho \sim \text{Unif}(0.5, 0.99)$, which perhaps could have been improved by incorporating domain knowledge. In any case, our analysis revealed a substantial amount of parameter uncertainty in this noisy, partially observed, low-population setting. We provide visualizations of the other parameter posterior estimates in the supplement.

For the second stage of analysis, we estimate the marginal likelihood of each model with normalizing flows (Algorithm~\ref{alg:nee}). The dimensionality of this dataset---a time series of length 59---posed a difficulty for our methodology. Normalizing flows, like all density estimation techniques, struggle with the curse of dimensionality. We reduced the dimensionality of this dataset and simulated trajectories via resampling, namely, computing the average incidence across distinct three-day windows. We show the smoothed version of the original data in Figure~\ref{fig:tdc-incidence}. This summary statistic reduces the dimensionality of the data from 59 to 20 while still capturing the long-term trends in the original outbreak, such as the sharp peak in week 1 and the smaller second wave that followed. As with our simulation experiments, we smoothed the discrete time series by adding a small amount of Gaussian noise to simulated trajectories. We ensured that random forest classifiers achieved no better than 60\% accuracy discriminating between ``real'' simulations and synthetic trajectories sampled from the trained normalizing flow.

Table~\ref{tab:hyp-testing-tdc} reports $p$-values resulting from the model misspecification test. Since we are comparing seven models and thus running seven tests on the same data, the family-wise error rate is 30\%, so we also list Benjamini-Hochberg adjusted $p$-values. After correcting, at a significance level of $0.05$, we fail to detect evidence of misspecification for any model. At the less conservative threshold of $\alpha=0.1,$ we would deem InH to be misspecified. InH's posterior predictions are misaligned with our observation, and the marginal posterior for $\rho$ is unusually concentrated around 1, which would accord with a failure to model the data well. Surprisingly, we fail to reject Base as misspecified at either level of $\alpha$, even though it provides no mechanism for reinfection and its posterior has the same problems as InH. The misspecification test appears underpowered for these data, perhaps due their the noise or dimensionality.

\begin{figure}
    \centering
    \includegraphics[width=0.5\linewidth]{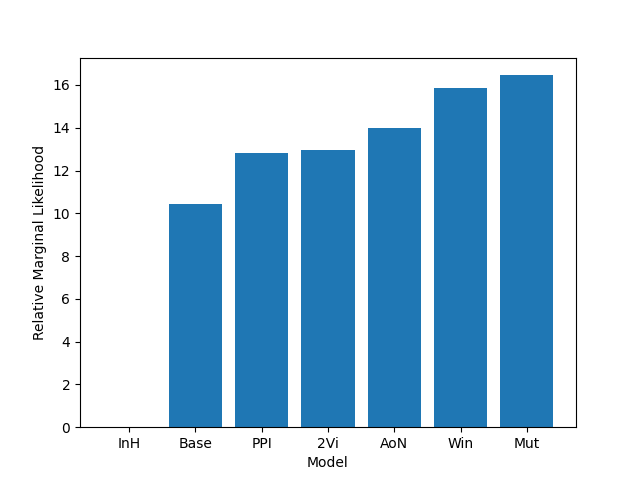}
    \caption{Log marginal likelihood of competing reinfection models, shown as a difference relative to the lowest-scoring model.}
    \label{fig:relative-marginal-likelihood}
\end{figure}

Figure~\ref{fig:relative-marginal-likelihood} visualizes the relative log marginal likelihoods of the seven models, ranked from least to greatest. For each model $\mdl_k$, $k=1, \ldots, 7$, the height of the bar is $\log p(\by_o \mid \mdl_k) - \log p(\by_o \mid \mdl_{l})$, where $\mdl_l$ is the model with the lowest evidence, in this case, InH. Thus, the height difference between any two bars is equivalent to the (log) Bayes factor between the two models. Our marginal likelihood analysis suggests that Win and Mut are the most plausible models of the 1971 TdC influenza epidemic. These models are substantially more likely than the 3rd place choice, AoN. PPI and 2Vi are virtually tied for the 4th likeliest. The Bayes factors imply that there is strong evidence against Base (and particularly) InH compared to the other five models.

Our analysis partially agrees with the findings of \textcite{camacho_explaining_2011}, who argued, by estimating the Akaike Information Criteria (AIC) for each model, that AoN and Win provided the best fit to the observed data, and that PPI and InH offered the worst fit, with the dual strain models Mut and and 2Vi ranked in the middle. The obvious difference between our results and theirs is that we found Mut to be the model with the greatest evidence.

AIC estimates the predictive error of the maximum likelihood estimate of a model's parameters, balancing a model's maximized log likelihood against a term penalizing the model's complexity:
\begin{equation}
    AIC = 2k - \log p(\bx_o \mid \bth = \boldsymbol{\hat \theta}),
\end{equation}
where $k$ is the number of model parameters and $ \boldsymbol{\hat \theta}$ is the maximum likelihood estimate \cite{cavanaugh2019akaike}. When comparing models with similar numbers of parameters, as in \textcite{camacho_explaining_2011}, AIC reduces to using the maximum log likelihood as the selection criterion. In contrast to AIC, the the model evidence takes parameter uncertainty into account by marginalizing the likelihood over the prior. This is a desirable property for noisy, low-population datasets such as the 1971 TdC influenza epidemic, where a wide range of parameter configurations may adequately explain the data (e.g. Figure~\ref{fig:rho_estimate}). In this instance, when there are significant model identifiability issues, the maximum likelihood point estimate may overfit the data and problematize AIC.

As they retain only Win and AoN as plausible models,\textcite{camacho_explaining_2011} conclude that host immunological heterogeneity is the main driver of reinfection and thus the multiple wave structure observed in the TdC epidemic. That is, a proportion of hosts have either a delayed or deficient humoral immune response to influenza infection \cite{camacho_does_2013}. Nonetheless, our results suggest that viral heterogeneity, particularly the evolution of a novel strain, may still play a role in the observed infection dynamics. Influenza strains that escape population immunity typically evolve over the course of months, not days, and we lack solid historical evidence of antigenic shift occurring in TdC. All seven models assume homogeneous mixing in the population, but it is possible that social mixing behaviors on the island varied day to day and contributed to changes in the virus's effective reproduction.

\begin{table}
\centering
\begin{tabular}{lrrr}
\toprule
Model & $p$-value & Corrected $p$ \\
\midrule
Base & \textbf{0.0425} & 0.149 \\
2Vi & 0.160 & 0.373  \\
Mut & 0.335 & 0.420  \\
AoN & 0.245 & 0.420 \\
PPI & 0.360 & 0.420  \\
InH & \textbf{0.0125} & \textbf{0.0875}  \\
Win & 0.443 & 0.443   \\
\bottomrule
\end{tabular}
\caption{Results of the misspecification detection test shown for competing reinfection models. The second column shows the raw $p$-values, and the third column shows their corrections via the Benjamini-Hochberg procedure. We bold $p$-values that lie below a significance threshold of $\alpha=0.1$.}
\label{tab:hyp-testing-tdc}
\end{table}

Finally, we use NEE to perform Bayesian model averaging. We generate predictions from an ensemble of the seven models weighted proportionally to their estimated evidence on the observed data (Equation~\ref{eq:bma}). The last row in Table~\ref{tab:tdc-errors} shows the MSPPE for the ensemble model across the three windows. Overall, the ensemble model yields an error slightly higher than the most accurate models. The ensemble's predictions during the second wave period (days 21-36) may explain this additional error. Examining the ensemble's posterior predictive distribution, shown in Figure~\ref{fig:tdc-predictive-checks}(h), we can see that the ensemble predicts a sharp spike in incidence in the first week and a reduction of new cases by day 40, as occurred in the observed data. However, the ensemble exhibits substantial volatility around the time of the second wave. That is, the ensemble predicts that secondary waves are likely, as opposed to models AoN, PPI, and Win, which have similar or lesser predictive error. We remark that minimizing in-sample mean squared error is not inherently a desirable property for a calibrated model, as this could result from overfitting observations. Similarly, ensembling models does not promise improved in-sample predictions but is rather considered to improve the robustness (i.e. generality) of predictions.

\section{Discussion}
\label{sec:discussion}

% worth noting that the marginal likelihood can depend heavily on the prior, even when the posterior doesn't (\cite{gelman}).

% cringe to talk about utility of AI for coding up many different models? 

% can at least mention that this method could have utility for ensemble modeling
 
Our experiments demonstrate that NEE is an effective, versatile, and simple tool for principled criticism of complex simulation models. When we know the true model generating our data, we find that NEE's estimates of model evidence are sufficiently accurate to reliably flag misspecified models and select the correct model. In contrast, predictive checks on their own are often insufficient for model comparison: they may favor overly simple or complex models in practice. Depending on the problem at hand, NEE can support hypothesis testing of model misspecification, model selection via Bayes factors, and Bayesian model averaging. NEE is a modular method that can be adapted to a range of simulators without significant tailoring. Its computational burden is light, since it avoids evaluation or approximation of potentially intractable model likelihoods. Training NEE on a modern MacBook Pro using a standard flow architecture and 2,000 training simulations per model, our networks typically converged in less than two minutes. 

% Our study has shown that NEE is an effective yet straightforward solution for model criticism in the setting of SBI avoiding several of the pitfalls of alternative methodologies (see also Section~\ref{sec:related}). By providing a measure of the degree of specification for one or more simulation models, NEE helps improve the robustness of NPE. Another attractive property of NEE is its light computational burden compared to older model criticism approaches. During experiments, we trained NEE on a modern MacBook Pro using a relatively simple flow architecture and 2,000 training simulations per model, and the networks typically converged in less than two minutes. NEE is a highly \textit{modular} methodology.

We nonetheless identify key limitations of NEE. First, NEE struggles to scale up to high-dimensional datasets. The curse of dimensionality is a ubiquitous challenge in statistical density estimation \cite{vandermeulen2024breaking}. Normalizing flows require an invertible architecture to transform inputs into a tractable target distribution, and so large dimensional datasets lead to poor convergence during training and daunting computational burdens. In epidemic modeling, data are often lengthy time series or outcomes stratified by a highly heterogeneous population. The use of summary statistics, such as working with final outcomes (Section~\ref{sec:experiment2}) or downsampling time series (Section~\ref{sec:data-analysis}) offers a partial workaround, but these can reduce the information content of the data and may elide crucial differences between competing models. 

% another idea: focus on modeling a smaller segment of the data
% cite derek hansen
Similarly, discrete data are problematic for NEE, since normalizing flows are built upon continuous variable transformations. Much of the data modeled by  biological simulations are discrete counts, e.g. the number of subjects currently in a compartment of a disease's progression. We added a small amount of Gaussian noise to smooth out count data, which empirically worked in simulation experiments but remains an \textit{ad hoc} fix (see also \textcite{hansen2022normalizing}). It is possible that the introduction of even a small amount of noise can bias estimates of the marginal likelihood. We note that discrete count data can hinder the sampling efficiency of widespread tools such as Iterated Filtering (IF) and PMCMC, which in practice add smoothing noise to simulations to assist parameter estimation. Future work on NEE for epidemic models could make use of flow architectures specialized for discrete data \cite{tran2019discrete} or incorporate a decreasing noise schedule during training, analogously to particle filtering.

Next, the reliability of model criticism with NEE depends on the accuracy of the neural estimates of the marginal likelihood.  We used Classifier Two-Sample tests---comparing true simulations from the prior predictive distribution with synthetic, flow-generated realizations via random forest classification---to judge whether our NEE estimators were ``accurate enough,'' but this test comes with some tolerance for approximation error \cite{lopez2016revisiting}.  Careful tuning of normalizing flows' hyperparameters may improve their convergence up to a point, as may increasing the size of the training set, though these incur additional computational and time costs.

The NEE algorithm estimates a model's evidence from a finite set of training simulations, so it tacitly assumes that the true marginal likelihood density for a model is reasonably regular, thereby making it possible to interpolate (or extrapolate) from the density of simulations to the density of observed data. While our simulation experiments suggest that NEE produces density estimates that correctly flag misspecification, this falls short of a theoretical guarantee bounding the error between NEE's predictions and the data's true marginal density. This is a limitation that NEE shares with NPE and other neural SBI techniques.
% also: enforcing a stricter agreement between simulations and synthetic samples?

During training, NEE learns network parameters $\psi^*$ that maximize the objective function
\begin{equation}
    \mathbf{E}_{p(\bx)} [\log h(\bx; \psi)],
\end{equation}
which translates to the finite sample approximation in Equation~\ref{eq:nee-emp}. Equivalently, NEE aims to minimize the forward KL divergence between the true marginal likelihood $p(\bx)$ and our estimator $h(\bx; \psi^*)$ \cite{ambrogioni2019forward}. The forward KL divergence is ``zero-avoiding,'' i.e. its minimizer places excess mass on low density regions of $p(\bx)$. (Intuitively, when $h(\bx; \psi)$ is close to zero, its logarithm rapidly goes to negative infinity, decreasing the above expectation.) NEE may therefore overestimate the evidence along the tail of its actual distribution, e.g. for misspecified models. This behavior would decrease the power of the model misspecification test and deflate Bayes factors. Nonetheless, under correct model specification, we found the the NEE $p$-values to be well calibrated with respect to the empirical false positive rate. This could mean that overestimation of the evidence may not matter for NEE as long as it accurately captures relative differences. 

Finally, identifiability issues impose a practical limit on the discriminatory power of NEE, though this is shared with other forms of model comparison. In Simulation Experiment I (Section~\ref{sec:experiment1}), the hypothesis test for model misspecification successfully diagnosed wrong models in less than half of all trials. While this may be explained at least in part by aforementioned errors in approximating the evidence, we surmised that many simulated datasets were ambiguous from the standpoint of model comparison, e.g. simulations where no outbreak occurred. Compared to the strict binary of a hypothesis test, Bayes factors may offer a potentially more nuanced mode of comparison, though they offer only a relative measure of evidence. Bayes factors may help with selecting a simpler model over a more complex model when neither are obviously misspecified, as in Simulation Experiment II. For the analysis of the 1971 TdC influenza epidemic, there was little clear-cut evidence pointing to the superiority of any one model. This could be explained by the noisiness of the observations and a lack of strong prior information on the incidence reporting rate, a weakly identifiable parameter. In situations such as this, we may consider a Bayesian average of all models under consideration to offer more robust predictions than any single candidate,
though ensembles sacrifice a degree of mechanistic interpretability.

\subsection{Conclusion}

% what is the frontier for infectious disease modeling?
While transmission models are an old idea, recent computational advances have helped make statistical modeling of epidemics an attainable goal. Just as NPE has shown promise for calibrating ambitious, large-scale simulations of infectious disease, we believe that NEE has the potential to advance rigorous multi-model analysis. In recent years, mechanistic transmission models applied to ongoing epidemics have proven useful for rapid decision-making \cite{desai2019real, zelner2022rapid, birrell2026real}. Though we undertook retrospective analysis of an epidemic as our case study, modeling epidemics in real time presents a possible direction for future work. Thanks to the computational speed of tools like NPE and NEE, a  simulation-based workflow may make it feasible to continuously calibrate and compare many complex transmission models as new observations come in.
 
NEE offers a practical means for measuring 
\textit{model uncertainty}, a critical yet sometimes misunderstood dimension of public health policy. During the COVID-19 pandemic, experts made inferences about policy recommendations based on many competing transmission models with differing assumptions. While these sometimes failed to yield accurate long-term forecasts, they still proved useful for projecting the effects of various public health interventions \cite{purkayastha2021comparison, keeling2022comparison}.
Transmission models are not oracles, and this should be communicated to the public. Thus, rigorously comparing the inferences of multiple informed transmission models is a more sensible goal than finding the ``one true model'' \cite{zelnerAccountingUncertaintyPandemic2021, reich2022collaborative}. 

Our work examined simulation-based criticism in the context of stochastic compartmental models of infectious disease, yet it can instruct future applications of simulation modeling. Transmission models have become familiar to the general public in recent years, as have the risks of taking an uncritical attitude to their predictions. They are illustrative of major challenges inherent to modeling partially observed, nonlinear dynamical systems. For similar reasons, many prominent models of genetic processes lack tractable likelihoods, so comparing them with conventional statistical tools is difficult (\cite{sullivan2005model, leuenberger2010bayesian}). Recently, NPE has demonstrated utility for efficiently calibrating complex models in population genetics and phylogenetics \cite{pinotti2026simulation, min2026neural}. Building on this work, NEE could support efficient model criticism in these fields. However, genetic data (e.g. site frequency spectra) are often have a large dimensionality, which presents a substantive challenge for flow-based estimation of model evidence. We anticipate that NEE and simulation-based criticism more broadly will improve the trustworthiness of neural SBI as it continues to find new applications.

\printbibliography

\appendix

\include{supplement1}

\end{document}

%% file: supplement1.tex
% \documentclass{article}

% \usepackage[utf8]{inputenc}
% \usepackage{booktabs, amsmath}
% \usepackage{graphicx}
% \begin{document}

\section{Appendix A: Details on SIR-like Compartmental Models}

We provide details on the stochastic compartmental models featured in the first simulation experiment (Section 3.1): Susceptible-Infected-Recovered (SIR), SIR with constant hazards (CH), Susceptible-Infected (SI), Susceptible-Infected-Recovered-Susceptible (SIRS), and Susceptible-Exposed-Infected-Recovered (SEIR).

\begin{table}[!htb]
\centering
\begin{tabular}{@{}llll@{}}
\toprule
Event            & Transition                & Rate                        & Model               \\ \midrule
Exposure         & $(S, E) \to (S - 1, E+1)$ & $\beta N^{-1}I(t) S(t)$ & SEIR                \\
Infection        & $(S, I) \to (S - 1, I+1)$ & $\beta N^{-1}I(t) S(t)$ & SIR, SI, SIRS,      \\
Infection        & $(S, I) \to (S - 1, I+1)$ & $\beta  S(t)$     & CH                  \\
Infection        & $(E, I) \to (E - 1, I+1)$ & $\sigma E(t)$               & SEIR                \\
Loss of Immunity & $(R, S) \to (R - 1, S+1)$ & $\delta R(t)$               & SIRS                \\
Recovery         & $(I, R) \to (I - 1, R+1)$ & $\gamma I(t)$               & SIR, CH, SIRS, SEIR \\ \bottomrule
\end{tabular}
\caption{Transition events for each of the SIR-like stochastic compartmental models. Though we implemented these models as discrete-time simulators, for ease of exposition we report instantaneous transition rates rather than binomial transition probabilities.}
\end{table}

\begin{table}[!htb]
\centering
\begin{tabular}{@{}lll@{}}
\toprule
Parameter & Prior                & Model               \\ \midrule
$\beta$   & $\text{Exp}(1)$ & SIR, SI, SIRS, SEIR \\
$\beta$   & $\text{Exp}(1/7)$ & CH                  \\
$\sigma$  & $\text{Exp}(1)$ & SEIR                \\
$\delta$  & $\text{Exp}(1)$ & SIRS                \\
$\gamma$  & $\text{Exp}(1)$ & SIR, CH, SIRS, SEIR
\end{tabular}
\caption{Prior distributions used to simulate from each model. Note that we use the shape/rate parametrization of the Gamma distribution.}
\end{table}

\section{Appendix B: Parameter Estimation for Influenza Reinfection Models}

For a detailed description of the mechanisms used in each reinfection model, please see Camacho et al. (2011). For each model (including our naive SEIR baseline), we report their parameters, the priors we used, and the NPE posterior mean \& 90\% credible interval defined by the $(0.05, 0.95)$ quantiles. In general, we used exponential priors for rate parameters and uniform priors for probability parameters. We set these with reference to the maximum likelihood estimates and confidence intervals obtained by Camacho et al. (2011). For initial conditions, i.e. the number of infected and immune individuals at the start of the observation period, we plug in their point estimates. 

\begin{table}[!htb]
\centering
\begin{tabular}{@{}lll@{}}
\toprule
Parameter  & Interpretation  & Prior                    \\ \midrule
$\beta$    & Infection Rate  & $\text{Exp}(0.2)$        \\
$\epsilon$ & Exposure Rate   & $\text{Exp}(2)$          \\
$\nu$      & Recovery Rate   & $\text{Exp}(2)$          \\
$\rho$     & Reporting Prob. & $\text{Unif}(0.5, 0.99)$
\end{tabular}
\caption{Priors for base SEIR model.}
\end{table}

\begin{table}[!htb]
\centering
\begin{tabular}{@{}lll@{}}
\toprule
Parameter  & Interpretation  & Prior                    \\ \midrule
$\beta_1$    & Infection Rate A & $\text{Exp}(0.3)$        \\ 
$\beta_2$ & Infection Rate B & $\text{Exp}(0.15)$ \\
$\epsilon$ & Exposure Rate   & $\text{Exp}(2)$          \\
$\nu$      & Recovery Rate   & $\text{Exp}(2)$          \\
$\gamma$ & Immunization Rate & $\text{Exp}(10)$ \\
$\rho$     & Reporting Prob. & $\text{Unif}(0.5, 0.99)$
\end{tabular}
\caption{Priors for 2Vi model.}
\end{table}

\begin{table}[!htb]
\centering
\begin{tabular}{@{}lll@{}}
\toprule
Parameter  & Interpretation  & Prior                    \\ \midrule
$\beta$    & Infection Rate & $\text{Exp}(0.2)$        \\ 
$\epsilon$ & Exposure Rate   & $\text{Exp}(2)$          \\
$\nu$      & Recovery Rate   & $\text{Exp}(2)$          \\
$\gamma$ & Immunization Rate & $\text{Exp}(10)$ \\
$\rho$     & Reporting Prob. & $\text{Unif}(0.5, 0.99)$ \\
$\sigma$ & Cross-Immunity & $\text{Unif}(0.1, 0.7)$ \\
$\mu$ & Mutation Time & $\text{Unif}(0, 59)$
\end{tabular}
\caption{Priors for Mut model.}
\end{table}

\begin{table}[!htb]
\centering
\begin{tabular}{@{}lll@{}}
\toprule
Parameter  & Interpretation  & Prior                    \\ \midrule
$\beta$    & Infection Rate & $\text{Exp}(0.2)$        \\ 
$\epsilon$ & Exposure Rate   & $\text{Exp}(2)$          \\
$\nu$      & Recovery Rate   & $\text{Exp}(2)$          \\
$\gamma$ & Immunization Rate & $\text{Exp}(10)$ \\
$\rho$     & Reporting Prob. & $\text{Unif}(0.5, 0.99)$ \\
$\alpha$ & Immunization Prob. & \text{Unif}(0.1, 0.9)
\end{tabular}
\caption{Priors for AoN model.}
\end{table}

\begin{table}[!htb]
\centering
\begin{tabular}{@{}lll@{}}
\toprule
Parameter  & Interpretation  & Prior                    \\ \midrule
$\beta$    & Infection Rate & $\text{Exp}(0.2)$        \\ 
$\epsilon$ & Exposure Rate   & $\text{Exp}(2)$          \\
$\nu$      & Recovery Rate   & $\text{Exp}(2)$          \\
$\gamma$ & Immunization Rate & $\text{Exp}(10)$ \\
$\rho$     & Reporting Prob. & $\text{Unif}(0.5, 0.99)$ \\
$\sigma$ & Protection Prob. & \text{Unif}(0.1, 0.7)
\end{tabular}
\caption{Priors for PPI model.}
\end{table}

\begin{table}[!htb]
\centering
\begin{tabular}{@{}lll@{}}
\toprule
Parameter  & Interpretation  & Prior                    \\ \midrule
$\beta$    & Infection Rate& $\text{Exp}(0.2)$        \\ 
$\epsilon$ & Exposure Rate   & $\text{Exp}(2)$          \\
$\nu$      & Recovery Rate   & $\text{Exp}(2)$          \\
$\gamma$ & Immunization Rate & $\text{Exp}(12)$ \\
$\rho$     & Reporting Prob. & $\text{Unif}(0.5, 0.99)$ \\
$\alpha$ & Clearance Prob. & \text{Unif}(0.1, 0.9)
\end{tabular}
\caption{Priors for InH model.}
\end{table}

\begin{table}[!htb]
\centering
\begin{tabular}{@{}lll@{}}
\toprule
Parameter  & Interpretation  & Prior                    \\ \midrule
$\beta$    & Infection Rate & $\text{Exp}(0.2)$        \\ 
$\epsilon$ & Exposure Rate   & $\text{Exp}(2)$          \\
$\nu$      & Recovery Rate   & $\text{Exp}(2)$          \\
$\gamma$ & Reinfection Window Rate & $\text{Exp}(10)$ \\
$\rho$     & Reporting Prob. & $\text{Unif}(0.5, 0.99)$ \\
$\tau$ & Immunization Rate. & \text{Unif}(0.1, 0.9)
\end{tabular}
\caption{Priors for Win model.}
\end{table}

\section{Appendix C: Miscellaneous Figures}

\begin{figure}[!htb]
    \centering
    \includegraphics[width=0.5\linewidth]{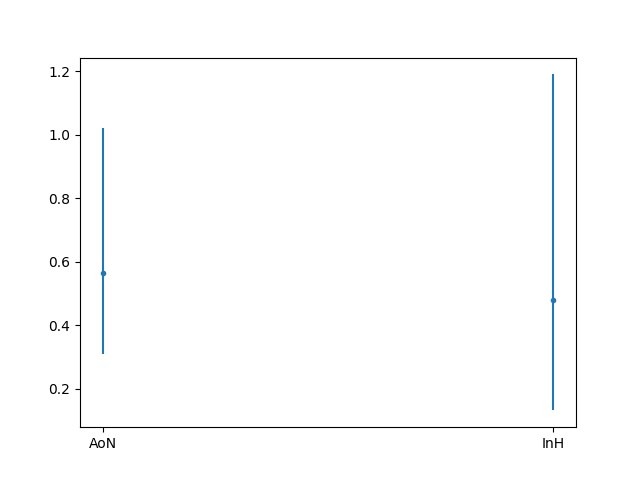}
        \caption{Posterior estimates of $\alpha$, showing the means and 90\% credible intervals.}
\end{figure}

\begin{figure}[!htb]
    \centering
    \includegraphics[width=0.5\linewidth]{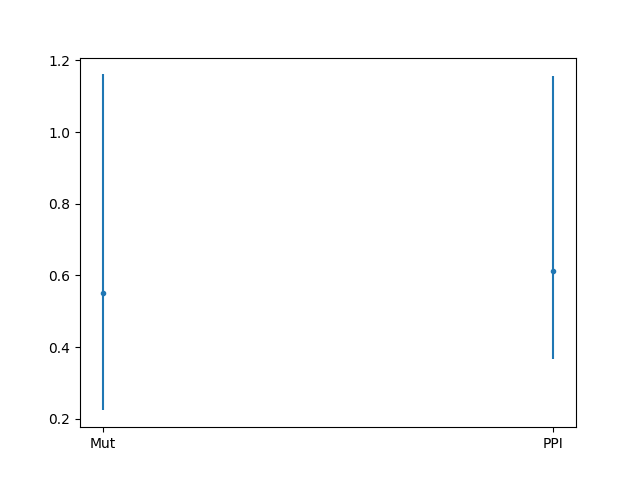}
        \caption{Posterior estimates of $1 - \sigma$, showing the means and 90\% credible intervals.}
\end{figure}

\begin{figure}[!htb]
    \centering
    \includegraphics[width=0.5\linewidth]{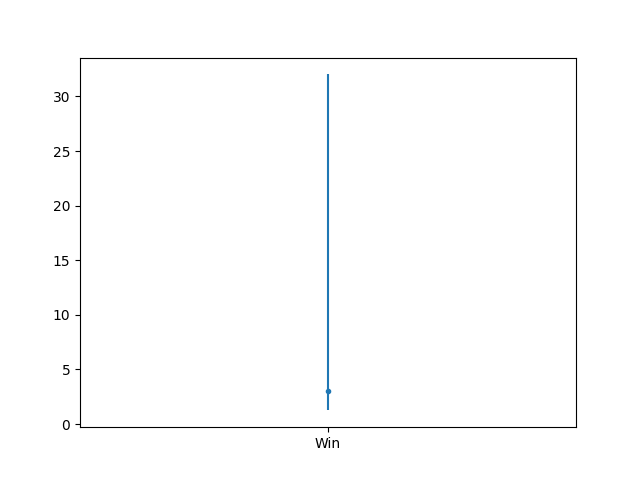}
        \caption{Posterior estimate of $1/ \tau $, showing the mean and 90\% credible interval.}
\end{figure}

\begin{figure}[!htb]
    \centering
    \includegraphics[width=0.5\linewidth]{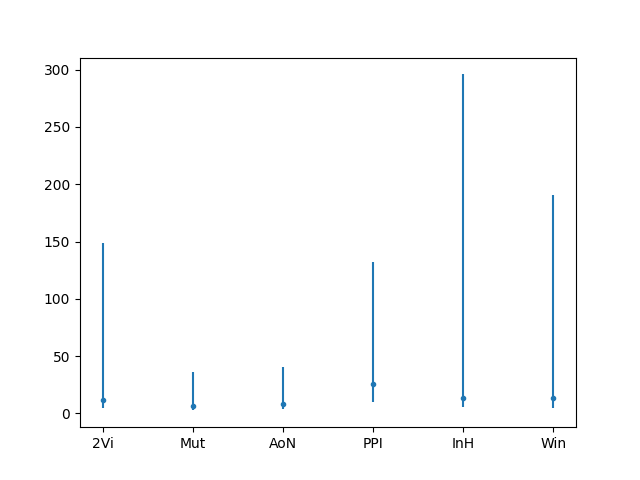}
        \caption{Posterior estimates of $1/ \gamma $, showing the means and 90\% credible intervals.}
\end{figure}

\begin{figure}[!htb]
    \centering
    \includegraphics[width=0.5\linewidth]{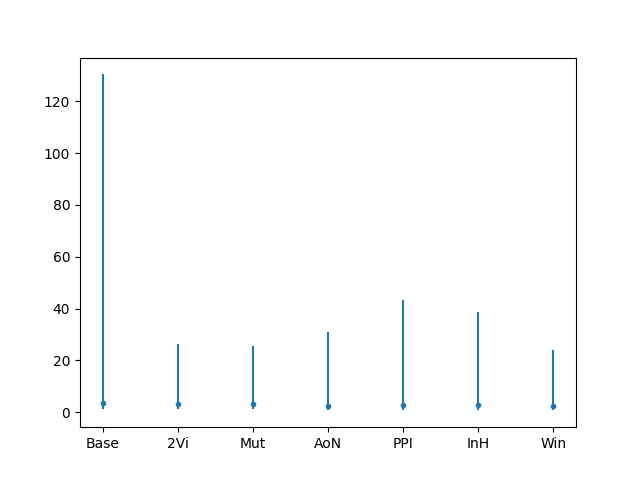}
        \caption{Posterior estimates of $1/ \nu $, showing the means and 90\% credible intervals.}
\end{figure}

\begin{figure}[!htb]
    \centering
    \includegraphics[width=0.5\linewidth]{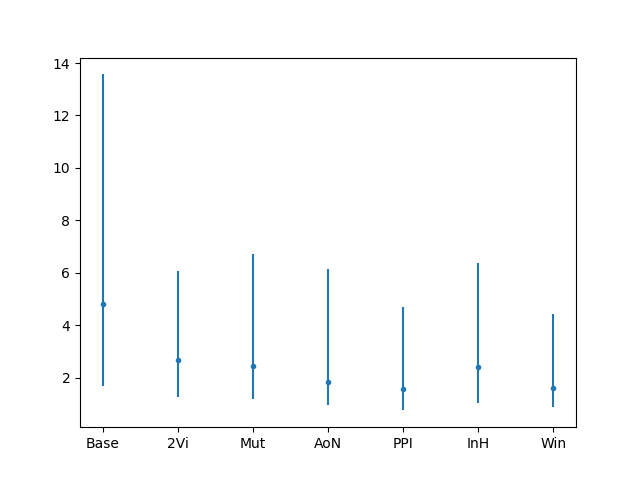}
        \caption{Posterior estimates of $1/ \epsilon $, showing the means and 90\% credible intervals.}
\end{figure}

\begin{figure}[!htb]
    \centering
    \includegraphics[width=0.5\linewidth]{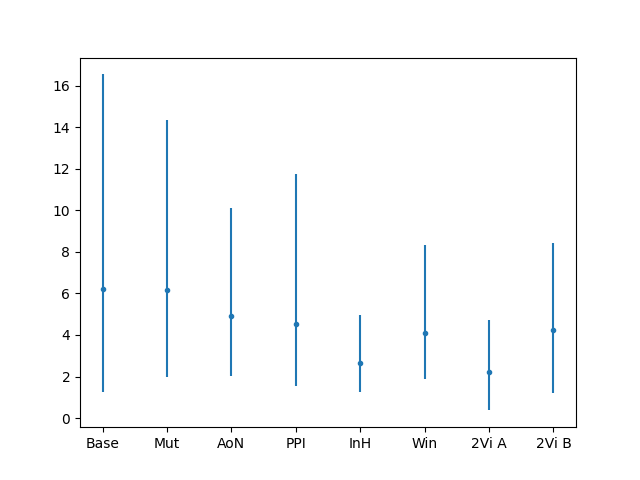}
        \caption{Posterior estimates of $\beta$, showing the means and 90\% credible intervals.}
\end{figure}

\begin{figure}[!htb]
    \centering
    \includegraphics[width=0.5\linewidth]{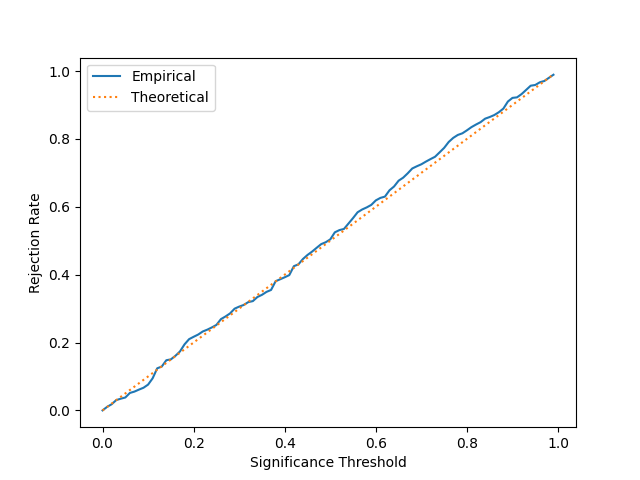}
    \caption{For 1,000 simulations of the SIR model (Simulation Experiment I), we show the false positive rate (FPR) of the model misspecification test, i.e. the number of trials in which we reject the null hypothesis that the SIR model is the true model. The close coincidence of the empirical and theoretical FPR indicates that the estimated $p$-values are generally well-calibrated.}
    \label{fig:fpr-calibration}
\end{figure}

% \end{document}

%% file: main.bib
@article{zelnerAccountingUncertaintyPandemic2021,
  title = {Accounting for uncertainty during a pandemic},
  author = {Zelner, Jon and Riou, Julien and Etzioni, Ruth and Gelman, Andrew},
  date = {2021-08-13},
  journaltitle = {Patterns},
  shortjournal = {Patterns (N Y)},
  volume = {2},
  number = {8},
  eprint = {34405155},
  eprinttype = {pubmed},
  pages = {100310},
  issn = {2666-3899},
  doi = {10.1016/j.patter.2021.100310},
  url = {https://pmc.ncbi.nlm.nih.gov/articles/PMC8361691/},
  urldate = {2026-02-18},
  pmcid = {PMC8361691}
}

@article{etherington2019mahalanobis,
  title={Mahalanobis distances and ecological niche modelling: correcting a chi-squared probability error},
  author={Etherington, Thomas R},
  journal={PeerJ},
  volume={7},
  pages={e6678},
  year={2019},
  publisher={PeerJ Inc.}
}

@misc{papamakarios_fast_2018,
	title = {Fast \${\textbackslash}epsilon\$-free Inference of Simulation Models with Bayesian Conditional Density Estimation},
	url = {http://arxiv.org/abs/1605.06376},
	number = {{arXiv}:1605.06376},
	publisher = {{arXiv}},
	author = {Papamakarios, George and Murray, Iain},
	urldate = {2024-02-16},
	date = {2018-04-02},
	langid = {english},
	eprinttype = {arxiv},
	eprint = {1605.06376 [cs, stat]},
}

@inproceedings{ambrogioni2019forward,
  title={Forward amortized inference for likelihood-free variational marginalization},
  author={Ambrogioni, Luca and G{\"u}{\c{c}}l{\"u}, Umut and Berezutskaya, Julia and Borne, Eva and G{\"u}{\c{c}}l{\"u}t{\"u}rk, Yaǧmur and Hinne, Max and Maris, Eric and Gerven, Marcel},
  booktitle={The 22nd International Conference on Artificial Intelligence and Statistics},
  pages={777--786},
  year={2019},
  organization={PMLR}
}

@article{toni_approximate_2009,
	title = {Approximate Bayesian computation scheme for parameter inference and model selection in dynamical systems},
	volume = {6},
	issn = {1742-5689, 1742-5662},
	url = {https://royalsocietypublishing.org/doi/10.1098/rsif.2008.0172},
	doi = {10.1098/rsif.2008.0172},
	pages = {187--202},
	number = {31},
	journaltitle = {Journal of The Royal Society Interface},
	shortjournal = {J. R. Soc. Interface.},
	author = {Toni, Tina and Welch, David and Strelkowa, Natalja and Ipsen, Andreas and Stumpf, Michael P.H},
	urldate = {2025-12-09},
	date = {2009-02-06},
	langid = {english},
}

@inproceedings{schmitt2023detecting,
  title={Detecting model misspecification in amortized Bayesian inference with neural networks},
  author={Schmitt, Marvin and B{\"u}rkner, Paul-Christian and K{\"o}the, Ullrich and Radev, Stefan T},
  booktitle={DAGM German Conference on Pattern Recognition},
  pages={541--557},
  year={2023},
  organization={Springer}
}

@misc{ward_robust_2022,
	title = {Robust Neural Posterior Estimation and Statistical Model Criticism},
	url = {http://arxiv.org/abs/2210.06564},
	doi = {10.48550/arXiv.2210.06564},
	number = {{arXiv}:2210.06564},
	publisher = {{arXiv}},
	author = {Ward, Daniel and Cannon, Patrick and Beaumont, Mark and Fasiolo, Matteo and Schmon, Sebastian M.},
	urldate = {2025-12-09},
	date = {2022-10-12},
	langid = {english},
	eprinttype = {arxiv},
	eprint = {2210.06564 [stat]},
}

@article{knock_bayesian_2014,
	title = {Bayesian model choice for epidemic models with two levels of mixing},
	volume = {15},
	issn = {1465-4644, 1468-4357},
	url = {https://academic.oup.com/biostatistics/article-lookup/doi/10.1093/biostatistics/kxt023},
	doi = {10.1093/biostatistics/kxt023},
	pages = {46--59},
	number = {1},
	journaltitle = {Biostatistics},
	shortjournal = {Biostatistics},
	author = {Knock, E. S. and O'Neill, P. D.},
	urldate = {2025-12-09},
	date = {2014-01-01},
	langid = {english},
}

@article{camacho_does_2013,
	title = {Does homologous reinfection drive multiple-wave influenza outbreaks? Accounting for immunodynamics in epidemiological models},
	volume = {5},
	issn = {17554365},
	url = {https://linkinghub.elsevier.com/retrieve/pii/S175543651300042X},
	doi = {10.1016/j.epidem.2013.09.003},
	shorttitle = {Does homologous reinfection drive multiple-wave influenza outbreaks?},
	pages = {187--196},
	number = {4},
	journaltitle = {Epidemics},
	shortjournal = {Epidemics},
	author = {Camacho, Anton and Cazelles, B.},
	urldate = {2025-12-09},
	date = {2013-12},
	langid = {english},
}

@article{hooten_guide_2015,
	title = {A guide to Bayesian model selection for ecologists},
	volume = {85},
	rights = {http://onlinelibrary.wiley.com/{termsAndConditions}\#vor},
	issn = {0012-9615, 1557-7015},
	url = {https://esajournals.onlinelibrary.wiley.com/doi/10.1890/14-0661.1},
	doi = {10.1890/14-0661.1},
	pages = {3--28},
	number = {1},
	journaltitle = {Ecological Monographs},
	shortjournal = {Ecological Monographs},
	author = {Hooten, M. B. and Hobbs, N. T.},
	urldate = {2025-12-09},
	date = {2015-02},
	langid = {english},
}

@article{camacho_explaining_2011,
	title = {Explaining rapid reinfections in multiple-wave influenza outbreaks: Tristan da Cunha 1971 epidemic as a case study},
	volume = {278},
	issn = {0962-8452, 1471-2954},
	url = {https://royalsocietypublishing.org/doi/10.1098/rspb.2011.0300},
	doi = {10.1098/rspb.2011.0300},
	shorttitle = {Explaining rapid reinfections in multiple-wave influenza outbreaks},
	pages = {3635--3643},
	number = {1725},
	journaltitle = {Proceedings of the Royal Society B: Biological Sciences},
	shortjournal = {Proc. R. Soc. B.},
	author = {Camacho, Anton and Ballesteros, Sébastien and Graham, Andrea L. and Carrat, Fabrice and Ratmann, Oliver and Cazelles, Bernard},
	urldate = {2025-12-09},
	date = {2011-12-22},
	langid = {english},
}

@article{spuriomancini_bayesian_2023,
	title = {Bayesian model comparison for simulation-based inference},
	volume = {2},
	rights = {https://creativecommons.org/licenses/by/4.0/},
	issn = {2752-8200},
	url = {https://academic.oup.com/rasti/article/2/1/710/7382245},
	doi = {10.1093/rasti/rzad051},
	pages = {710--722},
	number = {1},
	journaltitle = {{RAS} Techniques and Instruments},
	author = {Spurio Mancini, A and Docherty, M M and Price, M A and {McEwen}, J D},
	urldate = {2025-12-09},
	date = {2023-01-17},
	langid = {english},
}

@article{britton_stochastic_2010,
	title = {Stochastic epidemic models: A survey},
	volume = {225},
	issn = {0025-5564},
	url = {https://www.sciencedirect.com/science/article/pii/S0025556410000143},
	doi = {10.1016/j.mbs.2010.01.006},
	shorttitle = {Stochastic epidemic models},
	pages = {24--35},
	number = {1},
	journaltitle = {Mathematical Biosciences},
	shortjournal = {Mathematical Biosciences},
	author = {Britton, Tom},
	urldate = {2025-12-09},
	date = {2010-05-01},
}

@article{he2009plug,
  title={Plug-and-play inference for disease dynamics: measles in large and small populations as a case study},
  author={He, Daihai and Ionides, Edward L and King, Aaron A},
  journal={Journal of the Royal Society Interface},
  volume={7},
  number={43},
  pages={271},
  year={2009}
}

@article{cranmer_frontier_2020,
	title = {The frontier of simulation-based inference},
	volume = {117},
	url = {https://www.pnas.org/doi/10.1073/pnas.1912789117},
	doi = {10.1073/pnas.1912789117},
	pages = {30055--30062},
	number = {48},
	journaltitle = {Proceedings of the National Academy of Sciences},
	author = {Cranmer, Kyle and Brehmer, Johann and Louppe, Gilles},
	urldate = {2025-12-09},
	date = {2020-12},
	note = {Publisher: Proceedings of the National Academy of Sciences},
}

@article{ionides2006inference,
  title={Inference for nonlinear dynamical systems},
  author={Ionides, Edward L and Bret{\'o}, Carles and King, Aaron A},
  journal={Proceedings of the National Academy of Sciences},
  volume={103},
  number={49},
  pages={18438--18443},
  year={2006},
  publisher={National Academy of Sciences}
}

@article{zelner2022rapid,
  title={Rapid response modeling of SARS-CoV-2 transmission},
  author={Zelner, Jon and Eisenberg, Marisa},
  journal={Science},
  volume={376},
  number={6593},
  pages={579--580},
  year={2022},
  publisher={American Association for the Advancement of Science}
}

@article{breto2009time,
  title={Time series analysis via mechanistic models},
  author={Bret{\'o}, Carles and He, Daihai and Ionides, Edward L and King, Aaron A},
  journal={The Annals of Applied Statistics},
  pages={319--348},
  year={2009},
  publisher={JSTOR}
}

@article{endo2019introduction,
  title={Introduction to particle Markov-chain Monte Carlo for disease dynamics modellers},
  author={Endo, Akira and Van Leeuwen, Edwin and Baguelin, Marc},
  journal={Epidemics},
  volume={29},
  pages={100363},
  year={2019},
  publisher={Elsevier}
}

@article{McKinley2018Approximate,
  title = {Approximate Bayesian computation and simulation-based inference for complex stochastic epidemic models},
  volume = {33},
  ISSN = {0883-4237},
  number = {1},
  journal = {Statistical Science},
  publisher = {Institute of Mathematical Statistics},
  author = {McKinley,  Trevelyan J. and Vernon,  Ian and Andrianakis,  Ioannis and McCreesh,  Nicky and Oakley,  Jeremy E. and Nsubuga,  Rebecca N. and Goldstein,  Michael and White,  Richard G.},
  year = {2018},
  month = feb 
}

@article{papamakarios2021normalizing,
  title={Normalizing flows for probabilistic modeling and inference},
  author={Papamakarios, George and Nalisnick, Eric and Rezende, Danilo Jimenez and Mohamed, Shakir and Lakshminarayanan, Balaji},
  journal={Journal of Machine Learning Research},
  volume={22},
  number={57},
  pages={1--64},
  year={2021}
}

@article{mackay1992bayesian,
  title={Bayesian interpolation},
  author={MacKay, David JC},
  journal={Neural Computation},
  volume={4},
  number={3},
  pages={415--447},
  year={1992},
  publisher={MIT Press One Rogers Street, Cambridge, MA 02142-1209, USA journals-info~…}
}

@article{frazier2020model,
  title={Model misspecification in approximate Bayesian computation: consequences and diagnostics},
  author={Frazier, David T and Robert, Christian P and Rousseau, Judith},
  journal={Journal of the Royal Statistical Society Series B: Statistical Methodology},
  volume={82},
  number={2},
  pages={421--444},
  year={2020},
  publisher={Oxford University Press}
}

@article{kleijn2012bernstein,
  title={The Bernstein-von-Mises theorem under misspecification},
  author={Kleijn, Bas JK and Van der Vaart, Aad W},
  journal={The Electronic Journal of Statistics},
  volume={6},
  pages={354-381},
  year={2012}
}

@article{lopez2016revisiting,
  title={Revisiting classifier two-sample tests},
  author={Lopez-Paz, David and Oquab, Maxime},
  journal={arXiv preprint arXiv:1610.06545},
  year={2016}
}

@article{BoeltsDeistler_sbi_2025,
  doi = {10.21105/joss.07754},
  url = {https://doi.org/10.21105/joss.07754},
  year = {2025},
  publisher = {The Open Journal},
  volume = {10},
  number = {108},
  pages = {7754},
  author = {Jan Boelts and Michael Deistler and Manuel Gloeckler and Álvaro Tejero-Cantero and Jan-Matthis Lueckmann and Guy Moss and Peter Steinbach and Thomas Moreau and Fabio Muratore and Julia Linhart and Conor Durkan and Julius Vetter and Benjamin Kurt Miller and Maternus Herold and Abolfazl Ziaeemehr and Matthijs Pals and Theo Gruner and Sebastian Bischoff and Nastya Krouglova and Richard Gao and Janne K. Lappalainen and Bálint Mucsányi and Felix Pei and Auguste Schulz and Zinovia Stefanidi and Pedro Rodrigues and Cornelius Schröder and Faried Abu Zaid and Jonas Beck and Jaivardhan Kapoor and David S. Greenberg and Pedro J. Gonçalves and Jakob H. Macke},
  title = {sbi reloaded: a toolkit for simulation-based inference workflows},
  journal = {Journal of Open Source Software}
}

@article{papamakarios2017masked,
  title={Masked autoregressive flow for density estimation},
  author={Papamakarios, George and Pavlakou, Theo and Murray, Iain},
  journal={Advances in Neural Information Processing Systems},
  volume={30},
  year={2017}
}

@article{box1980sampling,
  title={Sampling and Bayes’ inference in scientific modelling and robustness},
  author={Box, George EP},
  journal={Journal of the Royal Statistical Society Series A: Statistics in Society},
  volume={143},
  number={4},
  pages={383--404},
  year={1980},
  publisher={Oxford University Press}
}

@article{kypraios2017tutorial,
  title={A tutorial introduction to Bayesian inference for stochastic epidemic models using Approximate Bayesian Computation},
  author={Kypraios, Theodore and Neal, Peter and Prangle, Dennis},
  journal={Mathematical Biosciences},
  volume={287},
  pages={42--53},
  year={2017},
  publisher={Elsevier}
}

@article{hansen2022normalizing,
  title={Normalizing flows for knockoff-free controlled feature selection},
  author={Hansen, Derek and Manzo, Brian and Regier, Jeffrey},
  journal={Advances in Neural Information Processing Systems},
  volume={35},
  pages={16125--16137},
  year={2022}
}

@article{tran2019discrete,
  title={Discrete flows: Invertible generative models of discrete data},
  author={Tran, Dustin and Vafa, Keyon and Agrawal, Kumar and Dinh, Laurent and Poole, Ben},
  journal={Advances in Neural Information Processing Systems},
  volume={32},
  year={2019}
}

@article{box1976science,
  title={Science and statistics},
  author={Box, George EP},
  journal={Journal of the American Statistical Association},
  volume={71},
  number={356},
  pages={791--799},
  year={1976},
  publisher={Taylor \& Francis}
}

@article{liu2024using,
  title={Using cross-validation methods to select time series models: Promises and pitfalls},
  author={Liu, Siwei and Zhou, Di Jody},
  journal={British Journal of Mathematical and Statistical Psychology},
  volume={77},
  number={2},
  pages={337--355},
  year={2024},
  publisher={Wiley Online Library}
}

@incollection{marin2018likelihood,
  title={Likelihood-free model choice},
  author={Marin, Jean-Michel and Pudlo, Pierre and Estoup, Arnaud and Robert, Christian},
  booktitle={Handbook of approximate Bayesian computation},
  pages={153--178},
  year={2018},
  publisher={Chapman and Hall/CRC}
}

@article{robert2011lack,
  title={Lack of confidence in approximate Bayesian computation model choice},
  author={Robert, Christian P and Cornuet, Jean-Marie and Marin, Jean-Michel and Pillai, Natesh S},
  journal={Proceedings of the National Academy of Sciences},
  volume={108},
  number={37},
  pages={15112--15117},
  year={2011},
  publisher={National Academy of Sciences}
}

@article{cannon2022investigating,
  title={Investigating the impact of model misspecification in neural simulation-based inference},
  author={Cannon, Patrick and Ward, Daniel and Schmon, Sebastian M},
  journal={arXiv preprint arXiv:2209.01845},
  year={2022}
}

@article{mattei2019parsimonious,
  title={A parsimonious tour of bayesian model uncertainty},
  author={Mattei, Pierre-Alexandre},
  journal={arXiv preprint arXiv:1902.05539},
  year={2019}
}

@article{newton1994approximate,
  title={Approximate Bayesian inference with the weighted likelihood bootstrap},
  author={Newton, Michael A and Raftery, Adrian E},
  journal={Journal of the Royal Statistical Society Series B: Statistical Methodology},
  volume={56},
  number={1},
  pages={3--26},
  year={1994},
  publisher={Oxford University Press}
}

@article{llorente2023marginal,
  title={Marginal likelihood computation for model selection and hypothesis testing: an extensive review},
  author={Llorente, Fernando and Martino, Luca and Delgado, David and Lopez-Santiago, Javier},
  journal={SIAM review},
  volume={65},
  number={1},
  pages={3--58},
  year={2023},
  publisher={SIAM}
}

@article{durkan2019neural,
  title={Neural spline flows},
  author={Durkan, Conor and Bekasov, Artur and Murray, Iain and Papamakarios, George},
  journal={Advances in Neural Information Processing Systems},
  volume={32},
  year={2019}
}

@article{gillespie1976general,
  title={A general method for numerically simulating the stochastic time evolution of coupled chemical reactions},
  author={Gillespie, Daniel T},
  journal={Journal of Computational Physics},
  volume={22},
  number={4},
  pages={403--434},
  year={1976},
  publisher={Elsevier}
}

@article{gelman2017prior,
  title={The prior can often only be understood in the context of the likelihood},
  author={Gelman, Andrew and Simpson, Daniel and Betancourt, Michael},
  journal={Entropy},
  volume={19},
  number={10},
  pages={555},
  year={2017},
  publisher={MDPI}
}

@article{streiner2011correction,
  title={Correction for multiple testing: is there a resolution?},
  author={Streiner, David L and Norman, Geoffrey R},
  journal={Chest},
  volume={140},
  number={1},
  pages={16--18},
  year={2011},
  publisher={Elsevier}
}

@article{ranganathan2016common,
  title={Common pitfalls in statistical analysis: the perils of multiple testing},
  author={Ranganathan, Priya and Pramesh, CS and Buyse, Marc},
  journal={Perspectives in Clinical Research},
  volume={7},
  number={2},
  pages={106},
  year={2016}
}

@book{jeffreys1998theory,
  title={The Theory of Probability},
  author={Jeffreys, Harold},
  year={1998},
  publisher={OuP Oxford}
}

@article{kass1995bayes,
  title={Bayes factors},
  author={Kass, Robert E and Raftery, Adrian E},
  journal={Journal of the American Statistical Association},
  volume={90},
  number={430},
  pages={773--795},
  year={1995},
  publisher={Taylor \& Francis}
}

@article{kermack1927contribution,
  title={A contribution to the mathematical theory of epidemics},
  author={Kermack, William Ogilvy and McKendrick, Anderson G},
  journal={Proceedings of the Royal Society of London. Series A, Containing Papers of a Mathematical and Physical Character},
  volume={115},
  number={772},
  pages={700--721},
  year={1927},
  publisher={The Royal Society London}
}

@article{morsomme2025exact,
  title={Exact Bayesian inference for fitting stochastic epidemic models to partially observed incidence data},
  author={Morsomme, Rapha{\"e}l and Xu, Jason},
  journal={The Annals of Applied Statistics},
  volume={19},
  number={3},
  pages={2279--2293},
  year={2025},
  publisher={Institute of Mathematical Statistics}
}

@article{bu2022likelihood,
  title={Likelihood-based inference for partially observed epidemics on dynamic networks},
  author={Bu, Fan and Aiello, Allison E and Xu, Jason and Volfovsky, Alexander},
  journal={Journal of the American Statistical Association},
  volume={117},
  number={537},
  pages={510--526},
  year={2022},
  publisher={Taylor \& Francis}
}

@article{barry2008cross,
  title={Cross-protection between successive waves of the 1918--1919 influenza pandemic: epidemiological evidence from US Army camps and from Britain},
  author={Barry, John M and Viboud, C{\'e}cile and Simonsen, Lone},
  journal={The Journal of Infectious Diseases},
  volume={198},
  number={10},
  pages={1427--1434},
  year={2008},
  publisher={The University of Chicago Press}
}

@article{rios2009qualitative,
  title={Qualitative analysis of the level of cross-protection between epidemic waves of the 1918--1919 influenza pandemic},
  author={Rios-Doria, D and Chowell, G},
  journal={Journal of Theoretical Biology},
  volume={261},
  number={4},
  pages={584--592},
  year={2009},
  publisher={Elsevier}
}

@article{andreasen1997dynamics,
  title={The dynamics of cocirculating influenza strains conferring partial cross-immunity},
  author={Andreasen, Viggo and Lin, Juan and Levin, Simon A},
  journal={Journal of Mathematical Biology},
  volume={35},
  number={7},
  pages={825--842},
  year={1997},
  publisher={Springer}
}

@article{casagrandi2006sirc,
  title={The SIRC model and influenza A},
  author={Casagrandi, Renato and Bolzoni, Luca and Levin, Simon A and Andreasen, Viggo},
  journal={Mathematical Biosciences},
  volume={200},
  number={2},
  pages={152--169},
  year={2006},
  publisher={Elsevier}
}

@article{miller2010incidence,
  title={Incidence of 2009 pandemic influenza A H1N1 infection in England: a cross-sectional serological study},
  author={Miller, Elizabeth and Hoschler, Katja and Hardelid, Pia and Stanford, Elaine and Andrews, Nick and Zambon, Maria},
  journal={The Lancet},
  volume={375},
  number={9720},
  pages={1100--1108},
  year={2010},
  publisher={Elsevier}
}

@article{mathews2007biological,
  title={A biological model for influenza transmission: pandemic planning implications of asymptomatic infection and immunity},
  author={Mathews, John D and McCaw, Christopher T and McVernon, Jodie and McBryde, Emma S and McCaw, James M},
  journal={PLoS One},
  volume={2},
  number={11},
  pages={e1220},
  year={2007},
  publisher={Public Library of Science San Francisco, USA}
}

@article{gomes2004infection,
  title={Infection, reinfection, and vaccination under suboptimal immune protection: epidemiological perspectives},
  author={Gomes, M Gabriela M and White, Lisa J and Medley, Graham F},
  journal={Journal of Theoretical Biology},
  volume={228},
  number={4},
  pages={539--549},
  year={2004},
  publisher={Elsevier}
}

@article{mathews2010prior,
  title={Prior immunity helps to explain wave-like behaviour of pandemic influenza in 1918-9},
  author={Mathews, John D and McBryde, Emma S and McVernon, Jodie and Pallaghy, Paul K and McCaw, James M},
  journal={BMC Infectious Diseases},
  volume={10},
  number={1},
  pages={128},
  year={2010},
  publisher={Springer}
}

@article{mummert2013perspective,
  title={A perspective on multiple waves of influenza pandemics},
  author={Mummert, Anna and Weiss, Howard and Long, Li-Ping and Amig{\'o}, Jos{\'e} M and Wan, Xiu-Feng},
  journal={PloS One},
  volume={8},
  number={4},
  pages={e60343},
  year={2013},
  publisher={Public Library of Science San Francisco, USA}
}

@article{sullivan2005model,
  title={Model selection in phylogenetics},
  author={Sullivan, Jack and Joyce, Paul},
  journal={Annual Review of Ecology, Evolution, and Systematics},
  volume={36},
  number={1},
  pages={445--466},
  year={2005},
  publisher={Annual Reviews}
}

@article{leuenberger2010bayesian,
  title={Bayesian computation and model selection without likelihoods},
  author={Leuenberger, Christoph and Wegmann, Daniel},
  journal={Genetics},
  volume={184},
  number={1},
  pages={243--252},
  year={2010},
  publisher={Oxford University Press}
}

@article{gelman2013philosophy,
  title={Philosophy and the practice of Bayesian statistics},
  author={Gelman, Andrew and Shalizi, Cosma Rohilla},
  journal={British Journal of Mathematical and Statistical Psychology},
  volume={66},
  number={1},
  pages={8--38},
  year={2013},
  publisher={Wiley Online Library}
}

@book{andersson2012stochastic,
  title={Stochastic epidemic models and their statistical analysis},
  author={Andersson, Hakan and Britton, Tom},
  year={2012},
  publisher={Springer Science \& Business Media}
}

@article{black2012stochastic,
  title={Stochastic formulation of ecological models and their applications},
  author={Black, Andrew J and McKane, Alan J},
  journal={Trends in ecology \& evolution},
  volume={27},
  number={6},
  pages={337--345},
  year={2012},
  publisher={Elsevier}
}

@article{pinotti2026simulation,
  title={Simulation-based inference of epidemiological and phylodynamic models via neural posterior estimation},
  author={Pinotti, Francesco and Th{\'e}z{\'e}, Julien and Bailly, Xavier and Fourni{\'e}, Guillaume},
  journal={Proceedings of the Royal Society B: Biological Sciences},
  volume={293},
  number={2077},
  year={2026},
  publisher={The Royal Society}
}

@article{kosma2023neural,
  title={Neural ordinary differential equations for modeling epidemic spreading},
  author={Kosma, Chrysoula and Nikolentzos, Giannis and Panagopoulos, George and Steyaert, Jean-Marc and Vazirgiannis, Michalis},
  journal={Transactions on Machine Learning Research},
  year={2023}
}

@article{la2020epidemiological,
  title={An epidemiological neural network exploiting dynamic graph structured data applied to the COVID-19 outbreak},
  author={La Gatta, Valerio and Moscato, Vincenzo and Postiglione, Marco and Sperli, Giancarlo},
  journal={IEEE Transactions on Big Data},
  volume={7},
  number={1},
  pages={45--55},
  year={2020},
  publisher={IEEE}
}

@article{ahmad2025can,
  title={Can neural networks estimate parameters in epidemiology models using real observed data? MJ Ahmad and K. G{\"u}nel},
  author={Ahmad, Muhammad Jalil and G{\"u}nel, Korhan},
  journal={Applied Intelligence},
  volume={55},
  number={2},
  pages={133},
  year={2025},
  publisher={Springer}
}

@article{zuchowski2024rarely,
  title={Rarely Just About One Model: Multi-Model and Multi-Methodology Reasoning in Epidemic Modeling},
  author={Zuchowski, Lena},
  journal={Philosophy of Medicine},
  volume={5},
  number={1},
  year={2024}
}

@article{craddock2025bayesian,
  title={A Bayesian modelling framework with model comparison for epidemics with super-spreading},
  author={Craddock, Hannah and Spencer, Simon EF and Didelot, Xavier},
  journal={Infectious Disease Modelling},
  year={2025},
  publisher={Elsevier}
}

@article{thoren2024model,
  title={Model uncertainty, the COVID-19 pandemic, and the science-policy interface},
  author={Thor{\'e}n, Henrik and Gerlee, Philip},
  journal={Royal Society Open Science},
  volume={11},
  number={2},
  pages={230803},
  year={2024}
}

@article{metcalf2017opportunities,
  title={Opportunities and challenges in modeling emerging infectious diseases},
  author={Metcalf, C Jessica E and Lessler, Justin},
  journal={Science},
  volume={357},
  number={6347},
  pages={149--152},
  year={2017},
  publisher={American Association for the Advancement of Science}
}

@article{crepey2022challenges,
  title={Challenges for mathematical epidemiological modelling},
  author={Crepey, Pascal and No{\"e}l, Harold and Alizon, Samuel},
  journal={Anaesthesia, critical care \& pain medicine},
  volume={41},
  number={2},
  pages={101053},
  year={2022}
}

@article{shandross2026multi,
  title={Multi-Model Ensembles in Infectious Disease and Public Health: Methods, Interpretation, and Implementation in R},
  author={Shandross, Li and Howerton, Emily and Contamin, Lucie and Hochheiser, Harry and Krystalli, Anna and Consortium of Infectious Disease Modeling Hubs and Reich, Nicholas G and Ray, Evan L and Castro Rivadeneira, Alvaro J and Contamin, Lucie and others},
  journal={Statistics in medicine},
  volume={45},
  number={1-2},
  pages={e70333},
  year={2026},
  publisher={Wiley Online Library}
}

@article{sherratt2023predictive,
  title={Predictive performance of multi-model ensemble forecasts of COVID-19 across European nations},
  author={Sherratt, Katharine and Gruson, Hugo and Grah, Rok and Johnson, Helen and Niehus, Rene and Prasse, Bastian and Sandmann, Frank and Deuschel, Jannik and Wolffram, Daniel and Abbott, Sam and others},
  journal={Elife},
  volume={12},
  pages={e81916},
  year={2023},
  publisher={eLife Sciences Publications, Ltd}
}

@article{cavanaugh2019akaike,
  title={The Akaike information criterion: Background, derivation, properties, application, interpretation, and refinements},
  author={Cavanaugh, Joseph E and Neath, Andrew A},
  journal={Wiley Interdisciplinary Reviews: Computational Statistics},
  volume={11},
  number={3},
  pages={e1460},
  year={2019},
  publisher={Wiley Online Library}
}

@article{bates2024cross,
  title={Cross-validation: what does it estimate and how well does it do it?},
  author={Bates, Stephen and Hastie, Trevor and Tibshirani, Robert},
  journal={Journal of the American Statistical Association},
  volume={119},
  number={546},
  pages={1434--1445},
  year={2024},
  publisher={Taylor \& Francis}
}

@misc{reich2022collaborative,
  title={Collaborative hubs: making the most of predictive epidemic modeling},
  author={Reich, Nicholas G and Lessler, Justin and Funk, Sebastian and Viboud, Cecile and Vespignani, Alessandro and Tibshirani, Ryan J and Shea, Katriona and Schienle, Melanie and Runge, Michael C and Rosenfeld, Roni and others},
  journal={American journal of public health},
  volume={112},
  number={6},
  pages={839--842},
  year={2022},
  publisher={American Public Health Association}
}

@article{cramer2022evaluation,
  title={Evaluation of individual and ensemble probabilistic forecasts of COVID-19 mortality in the United States},
  author={Cramer, Estee Y and Ray, Evan L and Lopez, Velma K and Bracher, Johannes and Brennen, Andrea and Castro Rivadeneira, Alvaro J and Gerding, Aaron and Gneiting, Tilmann and House, Katie H and Huang, Yuxin and others},
  journal={Proceedings of the National Academy of Sciences},
  volume={119},
  number={15},
  pages={e2113561119},
  year={2022},
  publisher={National Academy of Sciences}
}

@article{fragoso2018bayesian,
  title={Bayesian model averaging: A systematic review and conceptual classification},
  author={Fragoso, Tiago M and Bertoli, Wesley and Louzada, Francisco},
  journal={International Statistical Review},
  volume={86},
  number={1},
  pages={1--28},
  year={2018},
  publisher={Wiley Online Library}
}

@article{beaumont2002approximate,
  title={Approximate Bayesian computation in population genetics},
  author={Beaumont, Mark A and Zhang, Wenyang and Balding, David J},
  journal={Genetics},
  volume={162},
  number={4},
  pages={2025--2035},
  year={2002},
  publisher={Oxford University Press}
}

@incollection{greenwood2009stochastic,
  title={Stochastic epidemic modeling},
  author={Greenwood, Priscilla E and Gordillo, Luis F},
  booktitle={Mathematical and statistical estimation approaches in epidemiology},
  pages={31--52},
  year={2009},
  publisher={Springer}
}

@article{vandermeulen2024breaking,
  title={Breaking the curse of dimensionality in structured density estimation},
  author={Vandermeulen, Robert A and Tai, Wai Ming and Aragam, Bryon},
  journal={Advances in Neural Information Processing Systems},
  volume={37},
  pages={65852--65883},
  year={2024}
}

@article{keeling2022comparison,
  title={Comparison of the 2021 COVID-19 roadmap projections against public health data in England},
  author={Keeling, Matt J and Dyson, Louise and Tildesley, Michael J and Hill, Edward M and Moore, Samuel},
  journal={Nature communications},
  volume={13},
  number={1},
  pages={4924},
  year={2022},
  publisher={Nature Publishing Group UK, London}
}

@article{purkayastha2021comparison,
  title={A comparison of five epidemiological models for transmission of SARS-CoV-2 in India},
  author={Purkayastha, Soumik and Bhattacharyya, Rupam and Bhaduri, Ritwik and Kundu, Ritoban and Gu, Xuelin and Salvatore, Maxwell and Ray, Debashree and Mishra, Swapnil and Mukherjee, Bhramar},
  journal={BMC infectious diseases},
  volume={21},
  number={1},
  pages={533},
  year={2021},
  publisher={Springer}
}

@article{desai2019real,
  title={Real-time epidemic forecasting: challenges and opportunities},
  author={Desai, Angel N and Kraemer, Moritz UG and Bhatia, Sangeeta and Cori, Anne and Nouvellet, Pierre and Herringer, Mark and Cohn, Emily L and Carrion, Malwina and Brownstein, John S and Madoff, Lawrence C and others},
  journal={Health security},
  volume={17},
  number={4},
  pages={268--275},
  year={2019},
  publisher={SAGE Publications, Sage CA: Los Angeles, CA}
}

@article{birrell2026real,
  title={Real-time modelling of the SARS-CoV-2 pandemic in England 2020--2023: A challenging data integration},
  author={Birrell, Paul J and Blake, Joshua and Kandiah, Joel and Alexopoulos, Angelos and Van Leeuwen, Edwin and Pouwels, Koen B and Ghosh, Sanmitra and Starr, Colin and Walker, Ann Sarah and House, Thomas A and others},
  journal={Journal of the Royal Statistical Society Series A: Statistics in Society},
  volume={189},
  number={2},
  pages={898--922},
  year={2026},
  publisher={Oxford University Press, UK}
}

@article{min2026neural,
  title={Neural posterior estimation for population genetics},
  author={Min, Jiseon and Ning, Yuxin and Pope, Nathaniel S and Baumdicker, Franz and Kern, Andrew D},
  journal={Genetics},
  volume={233},
  number={3},
  pages={iyag107},
  year={2026},
  publisher={Oxford University Press, US}
}
